# Menopause averted a midlife energetic crisis with help from older dependent children and parents: A simulation study

Edward H. Hagen
Washington State University
edhagen@wsu.edu

**Abstract** **Objectives**: The grandmother hypothesis proposes that ancestral women ceased reproduction midlife to instead provision their grandchildren. An alternative "two-sex" account proposes that the high energetic burden of caring for slow-developing offspring was met with biparental investment. Menopause evolved because the physiological costs of reproduction increased with age, yet productivity also increased with age, and the benefits of resource transfers by parents and grandparents of both sexes to adult children and their offspring eventually outweighed the diminishing benefits of continued reproduction (Kaplan et al., 2010). The "father absent" hypothesis proposes that the higher mortality rate of husbands would often have left wives without the resources to raise young children, selecting for early reproductive cessation (Kuhle, 2007). Juvenile production plays little role in the three hypotheses, yet subsequent studies have found it to be surprisingly high. **Materials and methods**: Simulations were conducted of hunter-gatherer energy consumption and production across the lifespan, taking account of age- and sex-specific survivorship, interbirth intervals, and varying rates of foraging skill acquisition typical of contemporary foragers. **Results**: There is a pronounced midlife energy deficit that could be averted with the increasing production of maturing juveniles; midlife cessation of reproduction, which limited the number of mouths to feed; and energy transfers from older parents, and sometimes younger couples (e.g., brideservice). **Discussion**: Menopause emerges as an integral and necessary component of the unique human pattern of relatively short interbirth intervals, a long period of juvenile dependency, and extensive food sharing, supporting and extending the "two-sex" and grandmother hypotheses.

## 1 Introduction

Prolonged post-reproductive lifespans are rare in wild animals (Chapman et al., 2024; Ellis, Franks, Nattrass, Cant, et al., 2018; Monaghan & Ivimey-Cook, 2023; cf. Winkler & Goncalves, 2023), consistent with the expectation that somatic and reproductive functions should senesce at similar rates (Croft et al., 2015). The explanation for the oddly long post-reproductive lifespans of females in humans and a few other species, on the other hand, remains hotly debated (see Figure 1). In brief, byproduct or non-adaptive explanations include that these are artifacts of unusually benign environments (e.g., captivity, recent improvements in public health) that extend lifespan but not fertility; that these are epiphenomena of antagonistic pleiotropy; or that these are artifacts of selection for male longevity that extended female lifespan but not fertility.

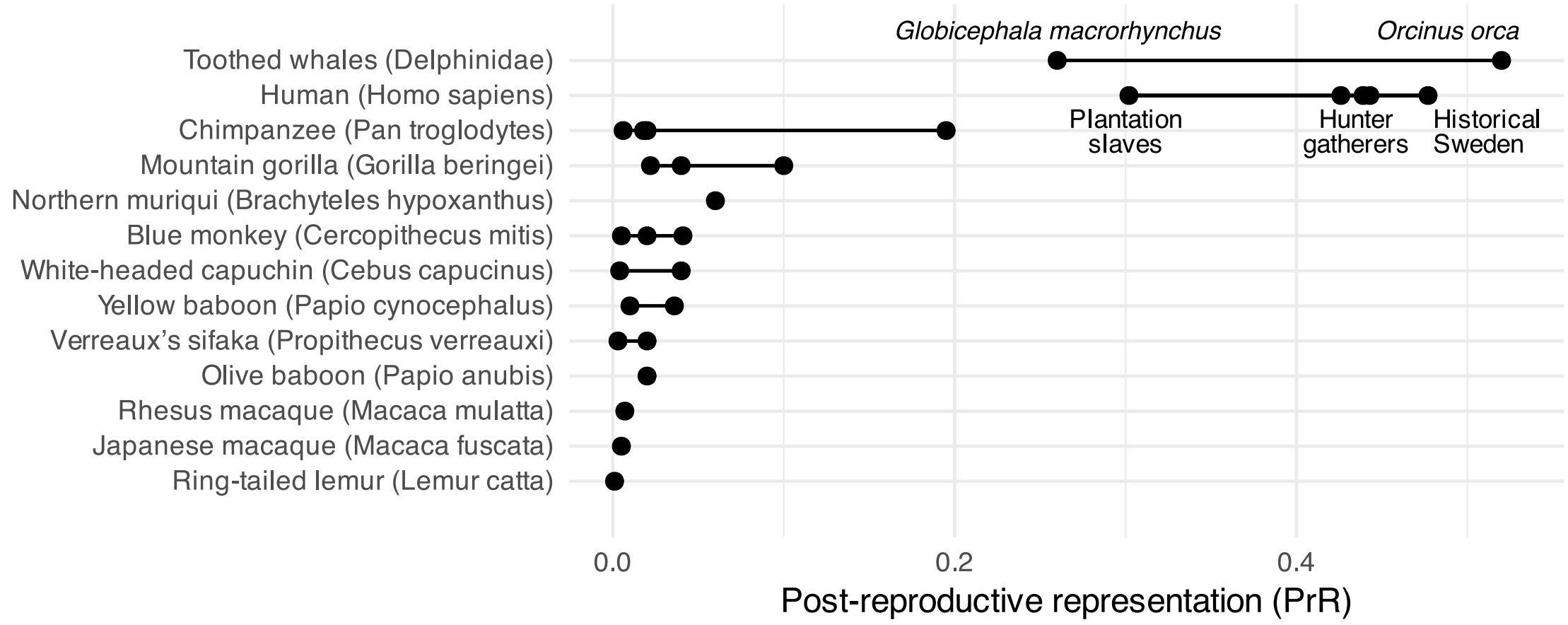

Figure 1: Post-reproductive lifespans of humans, African great apes and other primates, and toothed whales. Values represent the life expectancy from the end of the fertile period as a proportion of life expectancy from the beginning of the fertile period. Human and nonhuman primate values are from Wood et al. (2023), except for the largest gorilla value, which is from Smit & Robbins (2025). Toothed whale values are from Ellis, Franks, Nattrass, Currie, et al. (2018).

Adaptive explanations include the influential grandmother hypothesis, in which the benefits of investing in grandchildren outweighed the fitness benefits of producing more children; the mother hypothesis, in which the benefits of maintaining investment in multiple slow-developing offspring outweighed the increasing risk of maternal morbidity and mortality from continued reproduction; the "aging eggs" hypothesis, in which increased risk of chromosomal abnormalities selected for a cessation of reproduction; and the avoidance of breeding competition with genetic kin. These and other hypotheses have been extensively reviewed elsewhere (Arnot, 2021; Croft et al., 2015; Ellis et al., 2024; Hawkes, 2020, 2025; Holmes, 2019; Monaghan & Ivimey-Cook, 2023; Peccei, 2001; Sievert, 2024), so I will not revisit them here. Suffice it to say that debate continues regarding virtually all hypotheses, including between the class of byproduct hypotheses, such as hominin extension of lifespan but not fertility, vs. adaptive hypotheses, such as the grandmother hypothesis.

Here I investigate two adaptive hypotheses for the evolution of human menopause based on biparental care of offspring in light of new evidence that juvenile hunter-gatherers can produce a surprisingly large amount of food.

## 1.1 The "two-sex" embodied capital model of menopause

Kaplan et al. (2010) proposed a "two-sex" learning and skill-based model of the evolution of human menopause. In this account, which was inspired, in part, by research among the Tsimane, humans evolved to exploit a highly knowledge- and skill-based "niche" providing energy-rich but difficult to extract resources that required the rearing of slow-developing, large-brained, and thus energetically expensive offspring (Kaplan et al., 2000). Offspring were produced every few years but required almost two decades to become self-sufficient, placing a potentially high energetic burden on mothers, which, based on Tsimane data, would increase by delaying menopause. This burden is met by a co-evolving sexual division of labor in long-term monogamous marriages in which husbands specialized in skill-based acquisition of high-risk, high-return resources, with their skills and productivity increasing for about two decades beyond the age at which they became self-sufficient. Mothers specialized in equally skill-based child rearing and acquisition of lower-risk, lower-return resources (Kelly, 2013). Menopause evolved because the physiological costs of pregnancy and childbirth increased with age, and oocyte quality decreased with age, yet productivity also increased with age. In this scenario, around midlife the fitness benefits of downward intergenerational transfers of resources (food sharing) by both parents, whose acquisition was based on four decades of skill development, termed *embodied capital*, outweighed the fitness benefits of continued offspring production. In particular, Kaplan et al. (2010) provide empirical evidence that Tsimane grandfathers provide more calories to grandchildren than do grandmothers. This account, which could therefore be dubbed the *grandparents hypothesis*, is part of the highly influential *Embodied Capital Model* (ECM, Kaplan et al., 2000). See Figure 2A,B.

## 1.2 The absent father model of menopause

Kuhle (2007) proposed a two-part "absent father" hypothesis for the evolution of menopause in humans that assumes the evolutionary importance of a sexual division of labor in long-term marriages, similar to the scenario proposed by Kaplan and colleagues (Kaplan et al., 2010, 2000). The first part proposes that fathers who abandoned their wives for younger women would have left their ex-wives without "the necessary level of paternal investment to rear newborns to reproductive age" (p. 332). The second part proposes that because male mortality is greater than female mortality, and because husbands tend to be older than their wives, husbands would often have died before their wives, again leaving the wives without the resources necessary to rear newborns. Menopause would have limited the childrearing burden placed on older single mothers.

Although serial monogamy is not uncommon in the ethnographic record, Kaplan et al. (2010) emphasize the cost of mate-switching, and provide evidence that in small scale societies most children are produced and raised in long-term monogamous relationships. I will therefore set the wife abandonment hypothesis aside and focus on the "two-sex" aspect of the Kaplan et al. (2010) and Kuhle (2007) hypotheses.

## 1.3 New data on the ontogeny of foraging skills

The ECM, based on the data available at the time, emphasized that adult men produced a substantial surplus of energy from high return but difficult to acquire resources, with maximum production around age 35-40, reproductive-aged women produced approximately their own daily total energy expenditure (TEE), and post-reproductive women produced a modest surplus (Gurven et al., 2006; Kaplan et al., 2000; Walker et al., 2002). These surpluses supported a growing family of slowly developing offspring, each of whom would require energetic subsidies for up to 20 years. See Figure 2A,B.

Subsequently, two important sets of empirical results suggest a modification of the ECM might be necessary. First, in many populations, proficiency in hunting and other food production occurs earlier in adulthood, and is more constant throughout adulthood, than depicted for men in Figure 2A. A study of hunting based on ~23,000 records from more than 1800 individuals from 40 locations, for example — mostly men — found that skill and productivity increased more rapidly with age than envisioned in the ECM, peaking in the early 30′s, and remaining flatter across adulthood (Koster et al., 2020). See Figure 2E. A study of the ontogeny of a number of skills among Tsimane horticulturists found that most skills, including food production, had been acquired by age 20, although self- and peer-ratings of high proficiency in food production for both men and women increased into their 40′s, consistent with the ECM (Schniter et al., 2015). See Figure 2D.

Second, there is increasing evidence that younger children are able to acquire substantial calories from wild foods (Pretelli et al., 2024). Hadza children, for example, when followed during foraging trips (and not just measuring food returned to camp), often acquired more calories during the day than required by an adult (Crittenden et al., 2013; Froehle et al., 2019). See Figure 2C. A study of child foraging using published data from 28 societies found that their foraging returns, as a fraction of returns by young adults, increased approximately linearly with age, albeit more slowly for more skill-intensive, difficult-to-extract resources (tubers and game), but more rapidly for easier-to-extract resources (fruit and fish/shellfish) where adult levels of productivity were reached by adolescence (Pretelli et al., 2022).

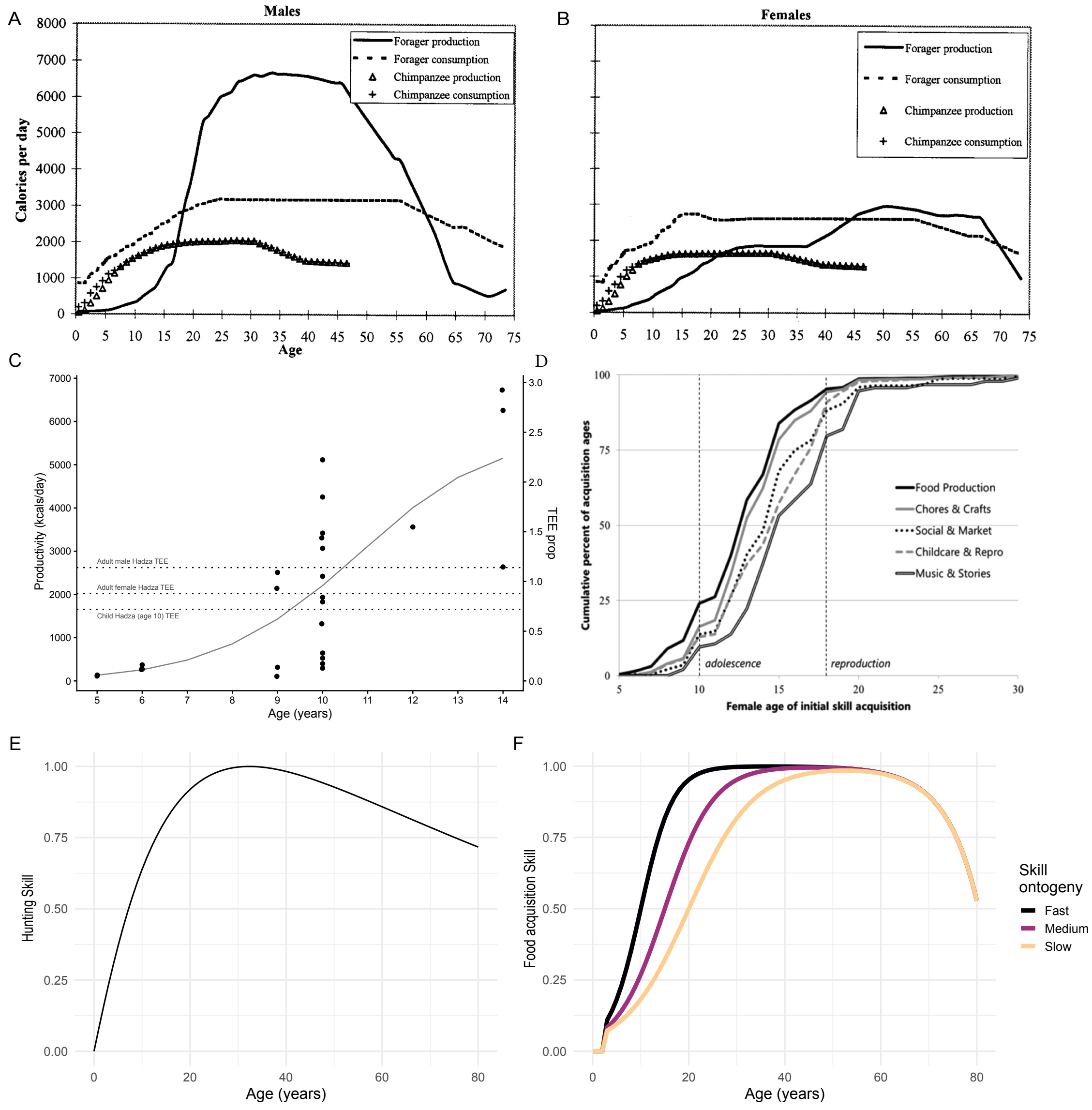

Figure 2: Age-specific energy production. **A, B**: Mean daily energy consumption and production values for male and female hunter-gatherers and chimpanzees that informed the original ECM. Figures from Kaplan et al. (2000). **C**: Hadza children daily productivity. Each dot is one foraging trip by one child. Right-hand y-axis is proportion of daily adult TEE. Data from Crittenden et al. (2013) and Froehle et al. (2019). **D**: Tsimane female skill acquisition (male skill acquisition is similar). Figure from Schniter et al. (2015). **E**: Age-specific hunting skill. Data and code from Koster et al. (2020). **F**: Age-specific food acquisition skill for the three values of $b_1$ and $age_{50}$ in Equation 3 that are used in the simulations. See also Table 1.

## 1.4 Study Aims

This study used simulations to investigate the energetic consequences of ages of last birth (ALB) ranging from much younger than typical of human hunter gatherers, to the median age for hunter-gatherers, to much closer to the end of life, resembling the pattern seen in other primates (Figure 1), taking into account the energetic consequences of a wider range of variation in foraging skill acquisition rates, juvenile energy production, and maximum adult energy production than in previous studies (Gurven & Walker, 2005; Kaplan et al., 2010). Other than ALB, all life history, demographic, and energetic factors fell within the ranges typical of contemporary hunter-gatherers.

Although this study was conceived, in part, as a test of the ECM "two-sex" hypothesis, the grandmother hypothesis similarly depends on female midlife reproductive cessation and postreproductive energy transfers. The implications of the results for both hypotheses will therefore be addressed.

It should be acknowledged at the outset that if menopause is an adaptation it co-evolved with demographic and life history factors, so the counterfactual that a species with, e.g., modern human interbirth intervals, maturation times, body masses, total energy expenditures, and lifespans, but much later menopause, or no menopause, is unrealistic. Nevertheless, this exercise could illuminate why it is unrealistic.

# 2 Methods

Taking inspiration, in part, from Gurven & Walker (2005), who evaluated the energetic consequences of slow vs. fast childhood growth, and from Kaplan et al. (2010), who evaluated family energy production and consumption using Tsimane data, energy production and consumption in hunter-gatherers was modeled in two ways.

## 2.1 The population model

The *population model* simulated the daily energy production and consumption of each age and sex class across the entire population, from birth to the maximum human lifespan (age 80), accounting for survivorship by age and sex. Energy consumption used age-, sex- and body-weight-adjusted total energy expenditure (TEE) values from the literature. Energy production was based on age- and sex-specific skills and physical strength (proxied by body weight), and varied across a range of parameter values for the relative importance of skill vs. strength, skill acquisition rates, and maximum adult energy production values (details below). Production, consumption, and their difference (energy balance) were computed for each age-sex class according to their proportions in a stationary population with a given life expectancy at birth and age- and sex-specific death rates given by a standard life table (Li & Gerland, 2012; Preston et al., 2001), and then summed for all ages and sexes. Positive values of energy balance would indicate that the population as a whole produces more energy than it consumes on a daily basis, and negative values that it consumes more energy than it produces.

This model was used to assess how variation in skill ontogeny and juvenile energy production impacted energy balance at the population level. It had the advantages that the proportions of each age and sex in the population corresponded to standard life tables (details in Section 2.5), energy transfers were inherent, and it made no assumptions about social organization – it was entirely agnostic regarding which individuals producing energy surpluses transferred them to which individuals in energy deficit. It had the disadvantages that it did not include variation in interbirth intervals or ages of last birth and their consequences for energy supply and demand.

## 2.2 The family model

Every individual has exactly one biological mother and biological father. To account for the energy production and consumption of everyone in the population, the *family model* simulated a hunter-gatherer nuclear family (mother, father, and their joint offspring) starting at age at first birth (AFB) and continuing with one-year time steps for a maximum human lifespan. The mother gave birth at a regular integer-valued interbirth interval (IBI) until her age at last birth (ALB), which was set to 30, 38, 50 and 60 years old. Juveniles matured and left the family at their AFB. Age- and sex-specific mortality followed standard human life tables (Li & Gerland, 2012). Energy production and consumption were computed as in the population model. Family energy balance was total family production minus total family consumption computed for each year from the AFB to the maximum age (80).

All members of the population resided in identical families that, for a given combination of parameters, followed identical trajectories across the lifecourse. A family with energy surpluses at some life stages (e.g., parents or grandparents) was assumed to transfer them to one family with energy deficits at other life stages (e.g., to the family of a young adult child). A negative sum of energy balance over the life of the family for a given combination of parameters would therefore indicate insufficient energy to support the families in the population, and thus a nonviable region of the parameter space.

This model was used to evaluate the impact of variation in ALB's, IBI's, and skill acquisition rates on energy balance. It had the advantage that it could vary ALB and IBI to determine their impact on family composition, and hence on energy balance. It had the disadvantage that varying ALB and IBI altered age-specific birth rates, violating population stationarity, as further discussed in Section 2.5.

This model conceptualized transfers of adult energy surpluses as biological parents feeding their spouses, joint offspring, and grandoffspring, per the ECM and absent father hypotheses, but the underlying computations only assumed that the energy surpluses produced by some individuals were used to compensate for the energy deficits of other individuals. In other words, the computations were agnostic regarding who transferred energy to whom. These transfers could therefore be equally conceptualized as, e.g., some men transferring their energy surpluses, not to a single wife and their joint offspring, but to any fertile woman and her offspring who are in energy deficit, in exchange for mating opportunities, per the grandmother hypothesis (Hawkes, 2020, 2025).[1]

## 2.3 Age- and sex-specific survivorship (mortality)

The average life expectancy at birth for human hunter-gatherers of both sexes combined is $e_0 = 30$ years (Davison & Gurven, 2021). For the Hadza, female $e_0 = 35.5$ and male $e_0 = 30.8$ (Blurton

[1]Due to the sex difference in mortality, the number of adult males in the simulations was always less than the number of adult females (Figure S 5), implying some degree of either polygyny or non-reproductive adult females. Simulations assumed all females reproduced, and the implied polygyny did not impact results, as further discussed in Section 4.2.

Jones, 2016). An analysis of !Kung mortality (Howell, 2017) found that it did not differ meaningfully from other human populations with similar life expectancy at birth (Coale & Demeny, 1966; Li & Gerland, 2012), a verdict echoed by Gurven & Kaplan (2007), who concluded that "there is a characteristic life span for our species, in which mortality decreases sharply from infancy through childhood, followed by a period in which mortality rates remain essentially constant to about age 40 years, after which mortality rises steadily in Gompertz fashion" (p. 322).

The absent father hypothesis (Kuhle, 2007) depends on a gap in the age of marriage (see Section 2.8) combined with a sex difference in mortality, which is common in many human populations as well as in many wild mammals. A study of 101 mammal species, for example, found that the median female lifespan was 18.6% longer than that of conspecfic males, and in humans the mean female advantage was 7.8% (Lemaître et al., 2020). For age- and sex-specific survivorship (mortality) I used the UN Model Life Tables (Li & Gerland, 2012; data and code from Gaddy et al., 2025) with $e_0$ values similar to the Hadza and !Kung, i.e., a female life expectancy at birth, $e_0 = 35$, a male life expectancy at birth, $e_0 = 30$, and a sex ratio at birth (SRB) of 1.05 (i.e., a 5% male excess). See Figure S 1.

## 2.4 Age at first birth (AFB), age at last birth (ALB), and interbirth interval (IBI)

Davison & Gurven (2021) compiled demographic and life history values for contemporary hunter-gather populations, small-scale subsistence populations, and chimpanzee populations. I primarily used their averaged values from the five hunter-gatherer populations: Ache, Agta, Hadza, Hiwi, and !Kung (Ju/'hoansi). Although Davison & Gurven (2021) did not provide a numerical value for mean AFB, their Figures 2 and S2 indicate hunter-gatherer values clustering from shortly before 20 years of age to shortly after. They found that the average AFB for all human subsistence populations was shortly before 20 years. Hunter-gatherer ALB ranged from 36 to 42, with a median of about 38, and mean total fertility was 6.2. Hunter-gatherer IBIs ranged from 2.8 years (Agta) to 3.3 years (Hiwi), with a mean value of 3.1 years. To match the foregoing values as closely as possible in the simulations, AFB = 20, IBI = 3 years and IBI = 4 years, and ALB = 38 was one of multiple values used.

A popular hypothesis for the evolution of menopause is that long-lived species can outlive their egg supply (Sievert, 2024). To determine the energetic consequences of earlier vs. later ALBs, it was therefore important to first determine the plausible range of ALBs beyond the hunter-gatherer median of 38. Females in most primates give birth until the end of life (Figure 1). A curvilinear regression model of the maximum ALB of mammals vs. maximum lifespan found that the predicted ALB of species with a maximum lifespan of 100 years, such as humans and narwals (*Monodon monoceros*), would be about 60 years (Huber & Fieder, 2018). The narwal ALB was 68 years, and three other species had ALB ≥ 60 years: African bush elephants (*Loxodonta africana*), Asian elephants (*Elephas maximus*), and North Atlantic right whales (*Eubalaena glacialis*). Moreover, this study noted that "there is little indication for reproductive senescence in any of the baleen whales so far", some of which have lifespans up to 200 years (Huber & Fieder, 2018, p. 2). Thus, because some mammals with maximum lifespans approaching or exceeding that of humans have ALB's approaching or exceeding 60 years, simulations were run with ALB = 30, 38, 50, and 60 years. For simplicity, in all simulation runs fertility was constant until ALB, at which point it was set to 0.

Birth at age x was conditional on female survival to age x. Taking into account that not all women survive to the ALB, in the simulations with IBI = 3, TFR = 6.1, and with IBI = 4, TFR = 4.5.

## 2.5 Population stationarity and growth

Contemporary foraging populations typically exhibit high annual population growth rates in excess of 1%, which is unsustainable over the long term, leading Gurven & Davison (2019) to propose that humans evolved as a colonizing species, with outmigration during periods of growth interspersed with relatively frequent population crashes. Populations are stationary, on the other hand, when births equal deaths over time (Preston et al., 2001). Among contemporary foragers, the !Kung were an exception to high population growth, with a relatively low rate of r = 0.17% attributable to a life expectancy at birth $e_0$ = 33.9, and a TFR = 4.3, close to the conditions for stationarity (Gurven & Davison, 2019). Roughly, with IBI = 4, and taking maternal mortality into account, mothers have about four children, about half of whom survive and reproduce, replacing their parents, yielding an approximately stationary population. The *population* and *family* models assume a stationary population with the age- and sex-specific mortality rate given by a standard life table (Li & Gerland, 2012; Preston et al., 2001). In the *family model*, each family's production supports itself and transfers any surplus to one family in energy deficit, e.g., to the family of one of its two surviving (or locally settling) adult offspring.

The Hadza had a more typical high growth rate of r = 1.38%, attributable to a similar $e_0$ = 34.7, but a higher TFR = 6.2 (Gurven & Davison, 2019). The simulations therefore used IBI values of 3 and 4 years, which, combined with the $e_0$ values specified in the previous section, approximate the TFR values (and hence the population growth rates) of the Hadza (high) and !Kung (low), respectively. In growing populations (IBI = 3), outmigration is assumed to keep the local population stationary.

## 2.6 Hunter-gatherer family energy consumption

Total energy expenditure (TEE) is a function of age, body mass, body composition (fat and fat-free mass), and activity levels. Age-related impacts on TEE include the energy required for development, age-related changes in activity levels, and tissue-specific metabolism. For example, the masses of energetically expensive organs, such as liver and brain, comprise a greater fraction of body mass in children than adults. A study of TEE measurements using doubly labeled water (the gold standard method) in a large, diverse sample found four distinct life stages: fat-free mass–adjusted expenditure accelerates rapidly from 0 to 1 year, declines slowly to adult levels by ~20 years, remains approximately stable from 20 to 60, even during pregnancy, and then declines in older adults (Bajunaid et al., 2025; Pontzer et al., 2021).

Somewhat counter-intuitively, after adjusting for age, sex, and body mass, TEE is relatively constant within these four phases, regardless of activity or immune activation. Instead, an individual's fixed energy budget is allocated among, e.g., growth, immunity, and activity (Pontzer et al., 2021). Children in horticultural Shuar vs. industrial populations have very similar TEE, for example, but compared to children in industrial populations, Shuar children have higher immune activation and activity, and reduced growth (Urlacher et al., 2019).

Accordingly, hunter-gatherer TEE was modeled using the regression model in Bajunaid et al. (2025), which comprises an equation that predicts TEE using age (years), sex, body mass (kg), height (cm), ethnicity, and elevation (m) (see Table 1 in Bajunaid et al., 2025), except that TEE for infants (age 0-1) was from Pontzer et al. (2021). Body mass and height were modeled as the average of generalized additive models (GAMs) of body masses and heights by age and sex of the Ache, Hadza, and !Kung. Ethnicity and elevation were set as "African" and 1000 m, respectively, yielding a mean adult female TEE = 2019 kcals/day, and a mean adult male TEE = 2581 kcals/day. See Figure S 2. Family energy consumption was the sum of TEE for the mother, father, and their dependent children for each age of the mother starting at AFB.

## 2.7 Adult hunter-gatherer energy production

A compilation of adult hunter-gatherer energy production by sex found that males typically (but not always) produced a surplus of calories beyond the typical adult male TEE, women generally produced around an adult female TEE (but in some cases with a surplus and in others with a deficit), and the combined male and female production was generally a surplus that would support offspring (Kraft et al., 2021). See Figure 3.

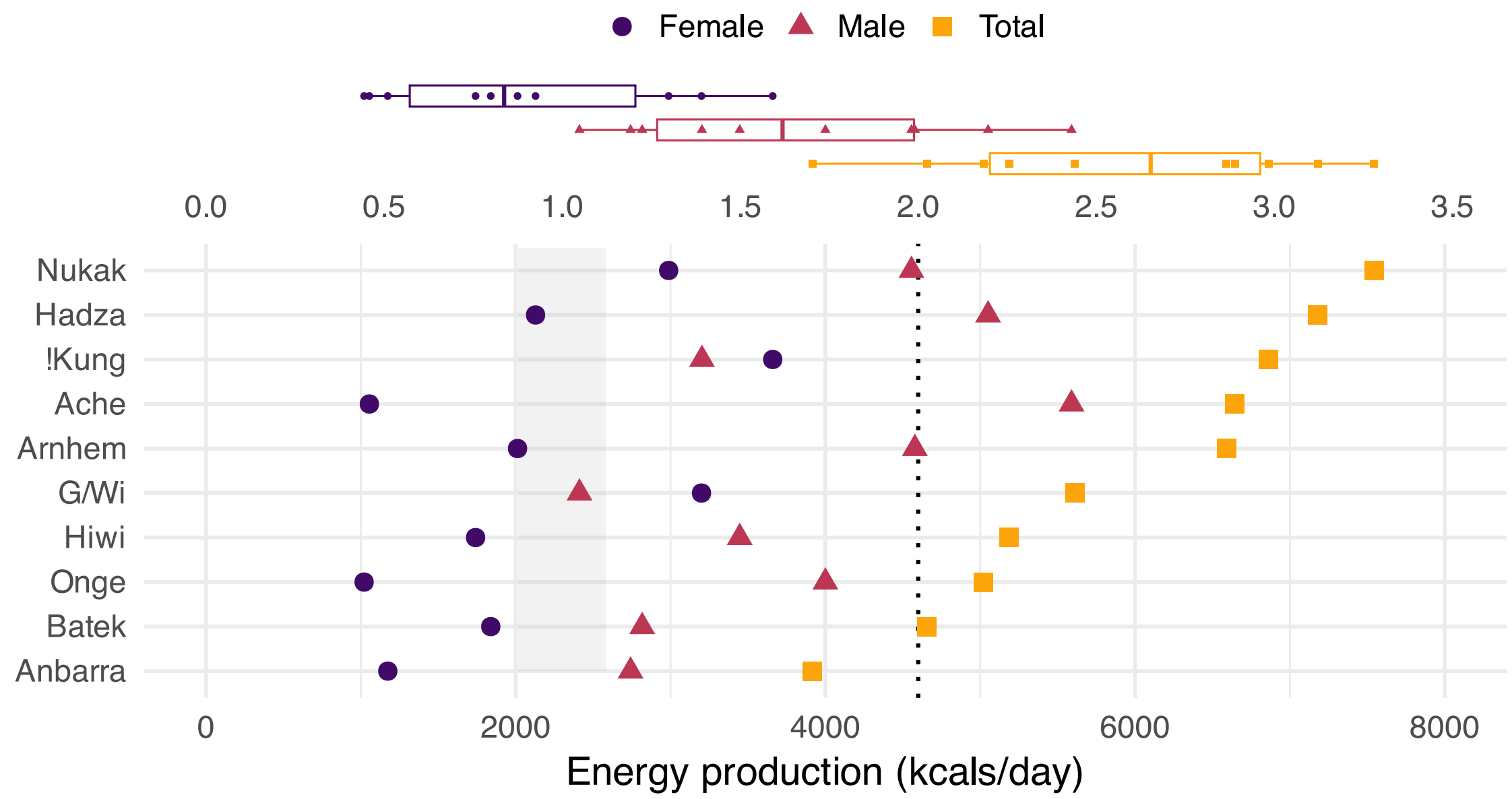

Figure 3: Empirical daily per capita adult energy production of contemporary hunter-gatherer females and males across ten populations. Boxplots indicate the distributions of female, male, and total production depicted in the main panel. Total: average production of one adult female plus one adult male. The grey bar indicates the range of an average female (left) to male (right) TEE for adult hunter-gatherers. The dotted line indicates total TEE for one female + male. Lower x-axis values are kcals/day. Upper x-axis values are kcals/day as a proportion of daily adult TEE ($TEE_{prop}$). Data from Kraft et al. (2021). Figure adapted from Venkataraman et al. (2026).

Taking the average TEE of a hunter-gatherer adult to be 2300 kcals/day, and using the energy production values in Figure 3 from Kraft et al. (2021), adult hunter-gatherer women produced from 0.4 to 1.6 of daily adult TEE, men from 1 to 2.4 of daily adult TEE, and one woman plus one man produced from 1.7 to 3.3 of daily adult TEE. Accordingly, maximum female energy production was modeled as a range from 0.4 to 1.6 of adult TEE, and maximum male energy production as a range from 1 to 2.2 of adult TEE, with the constraint that $2.2 \leq TEE_{prop,f} + TEE_{prop,m} \leq 3.4$ (i.e., the range needed to support at least two adults with some surplus for juvenile offspring, with a maximum slightly above the maximum joint production in Figure 3).

#### 2.7.1 Age- and sex-specific energy production

Hunter-gatherer energy production (kcals/day) by age and sex is fairly uncertain because there have been very few such populations to study, and measurements might be restricted to food returned to camp (ignoring food consumed during foraging), perhaps obtained only during parts of the year, and/or data from adults only. Drawing inspiration from modeling approaches in Gurven & Walker (2005), Gurven & Kaplan (2006), Koster et al. (2020), and Schniter et al. (2015), I modeled daily foraging productivity across the lifespan as follows:

$$productivity(age) = \text{TEE}_{prop} strength(age)^{\alpha} skill(age)^{1-\alpha} \tag{1}$$

where $TEE_{prop}$ is the proportion of adult TEE/day, and ranged from 0.4 to 1.6 for women and from 1 to 2.2 for men, and $0 \leq \alpha \leq 1$ is the relative importance of strength vs. skill, and was set to 0.25 (mostly skill-based), 0.5 (equal parts skill and strength), and 0.75 (mostly strength-based), separately for males and females. Strength was proxied by age- and sex-specific body weight (kg) as a proportion of maximum adult male weight, multiplied by a function that declines rapidly as age approaches the maximum age, representing senescence (e.g., reductions in muscle mass and bone density in old age):

$$strength(age, sex) = \frac{weight(age, sex)(1 - e^{b_0(age_{\max} - age)})}{weight_{\max}} \tag{2}$$

where $b_0$ is a parameter that determines the rate of strength decline in old age, and was fixed at −0.15 for both sexes in all simulation runs.

Skill is the product of a sigmoidal function that predominates at young ages, similar to that fit to the data in Schniter et al. (2015), with a function that declines rapidly as age approaches the maximum age:

$$skill(age) = \frac{1}{1 + e^{-b_1(age - age_{50})}}(1 - e^{b_0(age_{\max} - age)}) \tag{3}$$

where $b_1$ is a parameter that determines the rate of skill acquisition, and $age_{50}$ is the age at which skill is 50% of maximum. These two parameters were combined to create *Fast* ($b_1 = 0.4$, $age_{50} = 10$), *Medium* ($b_1 = 0.25$, $age_{50} = 15$) and *Slow* ($b_1 = 0.15$, $age_{50} = 20$) skill acquisition ontogenies, separately for each sex. $b_0$ is a parameter that determines the rate of skill decline in old age, and was fixed at −0.15 for both sexes in all simulation runs. Productivity was set to 0 for children less than 3 years old. See Figure 2F and Table 1.

Table 1: Ages at which food acquisition skill level reaches 50% and 95% of maximum value for each value of $b_1$ and $age_{50}$ used in the simulations. See Equation 3 and Figure 2F.

| Foraging skill ontogeny | $b_1$ | 50% | 95% |
|---|---|---|---|
| Fast | 0.40 | 10 | 17 |
| Medium | 0.25 | 15 | 27 |
| Slow | 0.15 | 20 | 40 |

## 2.8 Age gaps in marriage

The *absent father* hypothesis assumes that, on average, husbands were older than wives over the course of human evolution (Kuhle, 2007). Binford (2001) compiled ages at first marriage from the ethnographic record for men and women in 177 hunter-gatherer societies (see Figure S 3). The modal, median, and 75% quantile age differences were 2, 5, and 9.5 years, respectively. Gaps in ages at first marriage were assumed to represent age gaps within marriages. Marriage age gaps of 0, 2, 5, and 10 years were therefore used to compute the probability that the husband and the wife would be alive, and the number of dependents, for each age of the wife (for simplicity, age gaps were only used for this computation, and were not used in the simulations).

## 2.9 Parameter space and simulated lifecourse

In the *population model* of the hunter-gatherer lifecourse, one simulation was run for all 972 combinations of the skill ontogeny and energy production parameters in Table 2 (i.e., omitting IBI and ALB). In the *family model*, one simulation was run for all 7776 combinations of parameters, including IBI and ALB. In both models, the sum of maximum adult male and female energy production was constrained to $2.2 \leq TEE_{prop,f} + TEE_{prop,m} \leq 3.4$, as noted earlier.

In the *family model*, for each of the combinations of parameter values the following vectors were computed in sequence:

1. wife and husband ages were integer-valued vectors from AFB until the maximum age (80), at one-year intervals.
2. wife's probability of survival at each age relative to her AFB was computed with the female life table.
3. husband's probability of survival at each age relative to the wife's at AFB was computed with the male life table (thus, there was slightly less than one husband per wife at AFB).
4. pregnancies (coded as 0, 1) occurred every IBI years, starting at one year prior to AFB.
5. births (coded as 0, 1) occurred every IBI years, starting at AFB and ending at ALB, and then scaled by mother survival to age x.
6. child survival was age- and sex-specific survival multiplied by births (i.e., no new children in years with birth = 0).
7. wife, husband, and child TEE's at each age (as described earlier) were multiplied by their survival at each age.
8. wife, husband, and child productivities at each age were computed using Equation 1 and multiplied by their respective survival at each age.

9. The number, energy consumption, and energy production of all dependent children at each wife age were computed assuming offspring disperse (leave the family) at the AFB, and using a windowing function based on wife's age and wife survival (such that the number of children was scaled by the number of women who survived to give birth; see code).
10. family consumption and production at each wife age were the sums of wife's, husband's, and total dependent children's consumption and production, respectively.
11. family energy balance was family production minus family consumption for each age.
12. cumulative family energy balance was the cumulative sum of family energy balance at each age from AFB to the end of life (maximum age).

Neither family member survival, nor pregnancies, births, IBI's, or any other factor were contingent on energy balance. Simulations were deterministic, i.e., there were no stochastic parameters. See code for details.

Table 2: Parameter values for simulation runs. Each run was a unique combination of these values for each age of women from age at first birth (AFB) to the maximum age (80 years), with the constraint that $2.2 \le TEE_{prop,f} + TEE_{prop,m} \le 3.4$. Productivity is defined in Equation 1, strength in Equation 2, and skill in Equation 3.

| Parameter | Description | Values |
|---|---|---|
| $e_{0,f}$ | Female life expectancy at birth (years) | 35 |
| $e_{0,m}$ | Male life expectancy at birth (years) | 30 |
| $age_{max}$ | Maximum human age (years) | 80 |
| *SRB* | Sex ratio at birth | 1.05 |
| *AFB* | Age at first birth (years) | 20 |
| *ALB* | Age of last birth | 30, 38, 50, 60 |
| *IBI* | Interbirth interval (years) | 3, 4 |
| $age_{gap}$ | Marriage age gap (years) | 0, 2, 5, 10 |
| $\alpha_f$, $\alpha_m$ | Relative importance of skill vs. strength for females and males | 0.25, 0.5, 0.75 |
| $age_{50,f}$, $age_{50,m}$ | Age of 50% skill acquisition for females and males | 10, 15, 20 |
| $b_0$ | Rate of skill decline with increasing age | −0.15 |
| $b_{1,f}$, $b_{1,m}$ | Skill acquisition rate for females and males | 0.15, 0.25, 0.4 |
| $TEE_{prop,f}$ | Maximum productivity for females as a proportion of adult TEE | 0.4, 0.8, 1.2, 1.6 |
| $TEE_{prop,m}$ | Maximum productivity for males as a proportion of adult TEE | 1, 1.4, 1.8, 2.2 |

Simulations were conducted using R version 4.4.1 (2024-06-14). Code and data are available here: https://github.com/grasshoppermouse/menopause and here: https://doi.org/10.5281/zenodo.20817884.

# 3 Results

## 3.1 Population model

The main results of the *population model* demonstrate that juvenile production was essential for maintaining population-level energy balance near or above zero across most of the parameter space, and was unnecessary only at the highest level of adult production. See Figure 4. The mean effects of skill ontogeny parameter variation are depicted in Figure S 6. Variation in maximum adult energy production ($TEE_{prop}$) had by far the biggest effect, followed by the rate of skill acquisition ($b_1$) and the relative importance of strength vs. skill in energy production ($\alpha$). Dependents produced 60% of their TEE, on average, albeit with considerable variation across the parameter space. See Figure S 8.

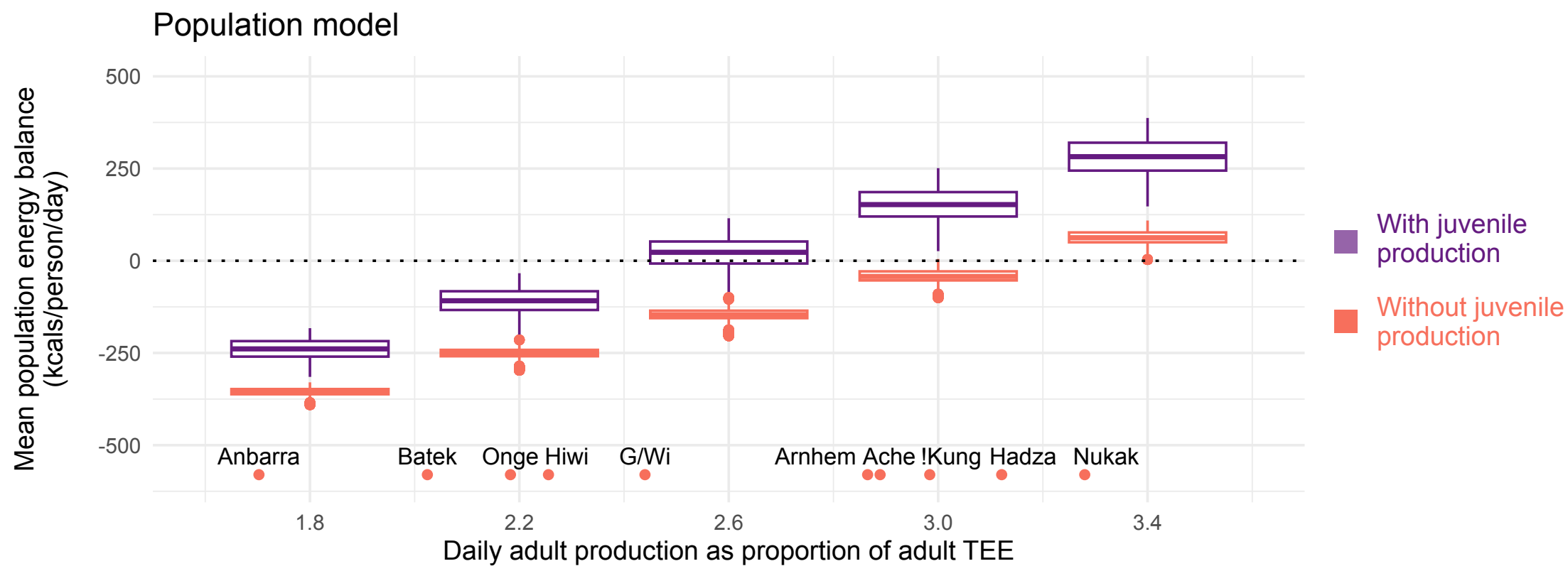

Figure 4: Mean per capita energy balance per day for different values of joint adult daily energy production (one adult female and male) as a proportion of adult TEE. Purple: energy balance including juvenile energy production. Orange: energy balance excluding juvenile production (age < 20). Boxplots represent variation in energy balance due to variation in skill ontogeny parameters. Hunter-gatherer populations along the x-axis are the joint adult energy production values from Figure 3, for comparison.

## 3.2 Family model

The main results of the *family model* demonstrate a tradeoff between women's total fertility and the capacity to feed her offspring. Averaging across the parameter space, as young couples produced offspring, family energy consumption tended to outpace production due to a short IBI relative to a long period of development, and families fell into negative energy balance. Although earlier reproductive cessation (ALB ≤ 38) reduced total fertility, it enabled families to eventually produce a surplus because their maturing dependent offspring produced an increasing fraction of their own energy needs (Figure S 7), and no new children were born. As mature offspring dispersed, energy surpluses could be transferred to families in energy deficit (e.g., to the new families of their newly dispersed adult offspring). With later reproductive cessation, however (ALB > 38), couples' families grew larger, and stayed large longer, and total daily energy needs were higher because women continued to give birth (see Table 3 and Figure S 9). Couples achieved higher total fertility but accumulated more energy "debt" and were less able to "pay it off" because they had fewer years later in the lifecourse to generate an energy surplus that could be transferred to families in energy deficit. See Figure 5.

Table 3: Demographic statistics by ALB and IBI, averaged over the parameter space, and across the lifespan from the wife's AFB (20) to her maximum age (80). Mean dependent consumption and production refer to totals of all dependent children. Mean energy balance includes the production and consumption of the wife, husband, and their dependents. Age values refer to the wife's age when the number of dependents reaches a maximum, when energy balance transitions from negative to positive, and when the number of dependents falls to zero.

| **Interbirth interval (IBI)** | **3** | **3** | **3** | **3** | **4** | **4** | **4** | **4** |
|---|---|---|---|---|---|---|---|---|
| **Age of last birth (ALB)** | **30** | **38** | **50** | **60** | **30** | **38** | **50** | **60** |
| Mean dependents | 0.8 | 1.3 | 1.9 | 2.2 | 0.6 | 1.0 | 1.4 | 1.7 |
| Maximum dependents | 2.8 | 4.2 | 4.2 | 4.2 | 2.2 | 3.1 | 3.1 | 3.1 |
| Mean family size | 1.9 | 2.4 | 3.0 | 3.3 | 1.7 | 2.0 | 2.5 | 2.8 |
| Maximum family size | 4.5 | 5.6 | 5.6 | 5.6 | 3.9 | 4.6 | 4.6 | 4.6 |
| Total fertility | 3.8 | 6.1 | 8.8 | 10.3 | 2.8 | 4.5 | 6.5 | 8.0 |
| Mean dependent consumption (kcals/day) | 1270 | 2077 | 2968 | 3487 | 960 | 1506 | 2194 | 2712 |
| Mean dependent production (kcals/day) | 856 | 1399 | 2000 | 2349 | 646 | 1015 | 1478 | 1827 |
| Mean family energy balance (kcals/day) | 225 | −40 | −332 | −503 | 327 | 148 | −78 | −248 |
| Age maximum dependents | 29 | 38 | 38 | 38 | 28 | 36 | 36 | 36 |
| Age positive energy balance | 36 | 45 | 57 | 68 | 34 | 41 | 53 | 67 |
| Age zero dependents | 49 | 58 | 70 | 79 | 48 | 56 | 68 | 80 |

Figure 5: Family model. **Top**: Mean family size by wife's age for different values of ALB (averaged across the different values of IBI). **Bottom**: Mean cumulative sum of daily family energy balance from the wife's AFB to the maximum lifespan, for different values of ALB (averaged across the parameter space). Values at age 80 near or above zero indicate that, on average, the family eventually produced enough total surplus energy to compenstate for its total energy deficits via energy transfers, whereas negative values indicate that the total surpluse fell short of the total deficit. Note the tradeoff between total fertility (averaged over IBI = 3 and IBI = 4) and energy balance for different values of ALB. Compare with the similar Figure 7 in Kaplan et al. (2010), and see also Figure 6. ALB: age of last birth. TF: total fertility.

Energy consumption, production, and balance tracked the number and ages of dependents, which were determined by the IBI and ALB. For a given IBI, family trajectories were identical prior to the youngest ALB (30 years), at which point they diverged. If there were no mortality, then for ALBs > 38, the number of dependents would remain constant until the ALB because as the oldest dependent child dispersed a new child would be born. In the simulations, however, the number of family members, including dependent children, slowly decreased due to husband mortality and to maternal mortality that also reduced the birth of new children.

Family energy consumption rose steady with the regular production of offspring whereas production increased slowly for about the first decade after AFB, resulting in early energy deficits (on average). Production increased more rapidly in the second decade as the productivity of maturing children increased, resulting in energy surpluses. Subsequently, energy consumption and production declined due to wife and husband mortality, and the departure of older productive children as they reached their AFB. See Figure 6.

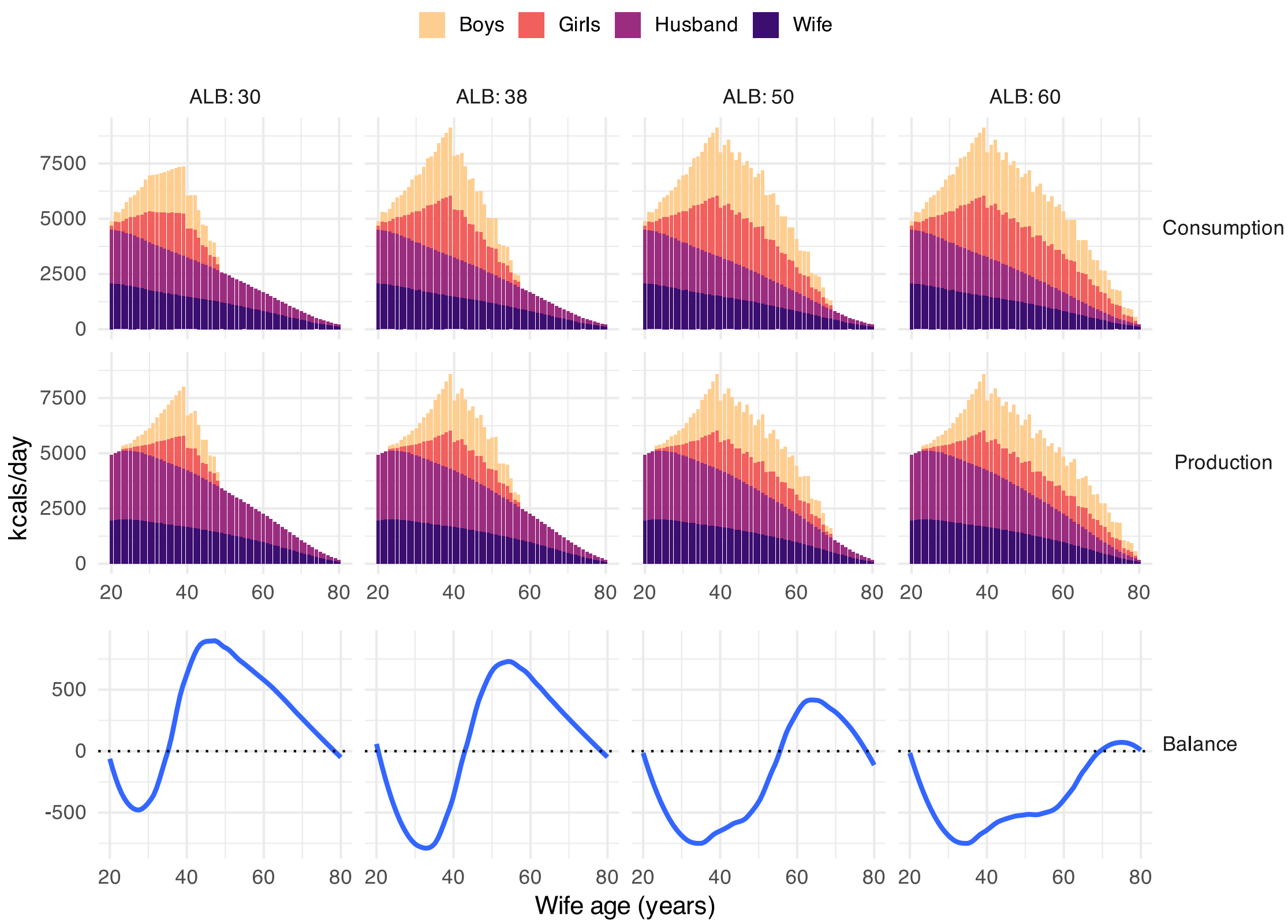

Figure 6: Daily energy consumption, production, and balance of the wife, husband, and their dependent children over the lifespan of the wife, averaged over the parameter space. Balance = production minus consumption. See Figure 5 for the cumulative sum of energy balance over the lifecourse.

Energy balance trajectories varied due to variation in the parameters affecting productivity, primarily $TEE_{prop}$ and $b_1$, each of which varied by sex. See Figure S 10. For $TEE_{prop} \leq 3$, which characterized all but two hunter-gatherer populations in Figure 3, families always experienced at least some time in negative energy balance across the parameter space; in other words, there was no combination of parameters in which families were entirely self-sufficient. At higher levels, however, there was a sharp transition: for $TEE_{prop} = 3.1$ (the level of Hadza joint energy production), families were entirely self-sufficient in just 5% of the parameter space, but for $TEE_{prop} = 3.3$ (the level of Nukak joint energy production , the highest in Figure 3), families were self-sufficient in 64% of the parameter space. See Figure S 11.

### 3.3 Brideservice and helpers-at-the-nest

Although in the simulations men and women both married at AFB, the age of marriage is often later for men than women. Whereas young women (age 20-25) produced a surplus in 52% of the parameter space, young men produced a surplus in 76% of the parameter space. See Figure S 12. In hunter-gatherer societies in which young adults remained with their natal families prior to marriage, net energy producers were serving as "helpers-at-the-nest" (Hagen & Barrett, 2009; Hames & Draper, 2004), whereas in those in which the bridegroom worked for the bride's family in exchange for marriage, termed brideservice (Walker et al., 2011), the surplus production of young men was helping provision their wives' younger siblings. Young couples with few or no offspring would also often produce a surplus that could provision, e.g., younger siblings.

### 3.4 Parent mortality

Mortality of both wives and husbands was high. With no marriage age gap, by age 50 only one parent had survived (the wife 54% of the time, the husband 45% of the time). At this age with IBI = 3 and ALB = 38, the single parent would be caring for 1.21 older dependent children, on average, whereas with ALB = 60, it would be 2.61 dependents, some probably quite young (e.g., 2, 5, and 8). With a marriage age gap of 10 years and ALB = 38, the single parent was somewhat more likely to be a younger wife, age 44 (the wife 62% of the time, the husband 38% of the time), who would be caring for 2.22 older dependents, on average, whereas with ALB = 60, it would be 2.97 dependents. See Figure S 13 and Figure S 14.

## 4 Discussion

The key insight from the simulations is that with a counterfactual human ALB closer to the end of life, similar to other primates (Figure 1), and given the energy consumption and production rates of contemporary foragers, the modern human pattern of relatively short interbirth intervals, long periods of juvenile dependency, minimal female reproductive skew, and long lifespans, was unlikely to have evolved. Shortly after AFB in most regions of the parameter space, younger parents with multiple highly dependent offspring would not be able to feed them without subsidies. Yet with a late ALB, older parents would still be burdened with many dependent children and would therefore be unable to subsidize younger parents in energy deficit – the population would be unable to feed everyone (Figure 5 and Figure 6). These results, which are based on parameter values from a wide range of hunter-gatherers, provide further support for the two-sex ECM model (Kaplan et al., 2010), which was developed using Tsimane data that indicated the importance of intergenerational energy transfers and the negative energetic impact of delayed menopause (see Figure S 4).

It is possible that, with a late ALB, ancestral foragers could simply have extracted more energy from the environment, rendering food sharing unnecessary. When the maximum joint production of wife and husband exceeded the needs of 3.2 adults ($TEE_{prop}$ > 3.2), which was greater than all contemporary hunter-gatherers in Figure 3 except the Nukak, families were indeed self-sufficient across much of the parameter space (Figure S 11). Under these conditions, however, families typically produced an energy surplus across the entire lifespan, implying that changes to human life history parameters would have evolved to take advantage of the surplus, such as changes to interbirth intervals, growth and maturation rates, and ages of first birth and dispersal.

The simulations therefore suggest the following role of menopause in the co-evolution of modern human life history traits. Relatively short interbirth intervals combined with long periods of juvenile dependence sow the seeds of a mid-life energetic crisis: parents have produced many young offspring with high energy requirements but who, per the ECM, require many years to acquire the skills necessary to provision themselves (see also Gurven & Walker, 2005). This crisis was averted with three evolved strategies.

First, by ceasing the production of new offspring in midlife, around age 40, parents limited the number of mouths to feed. Studies of contemporary humans and other species under resource constraints consistently show a quantity-quality tradeoff – offspring in larger families suffer deficits of various sorts (Gillespie et al., 2008; Hagen et al., 2006; Lawson & Mace, 2011; Walker et al., 2007). By limiting energetic investments to fewer slow-developing juveniles, parents enabled them to acquire the skills and strength necessary to support themselves before their dispersal, at which point parents began climbing out of an energy deficit even without subsidies (Figure 6). This refocuses the mother hypothesis on the energy production capacity of the mother and other family members.

Second, the increasing energy production of maturing juveniles, which tends to be overlooked in most theories of the evolution of menopause and other life history traits (Pretelli et al., 2024), was critical to meeting family energy needs (Figure 4). Even if juveniles could not completely support themselves, the combined production of multiple juveniles substantially reduced the total demand on parents. Furthermore, young adults, who had few or no children, often produced a surplus (Figure S 12) that could subsidize their immature siblings to whom they are as closely genetically related as to their own offspring, akin to the "helper at the nest" pattern seen in some other cooperatively breeding species (Hagen & Barrett, 2009; Hatchwell, 2009). This is also consistent with the brideservice practices of many hunter-gatherer societies in which the bridegroom works for the family of the bride, likely helping feed her younger siblings, in exchange for marriage (brideprice is a related practice in which the bridegroom's family pays the bride's family with some form of wealth, Goody & Tambiah, 1973; Walker et al., 2011).

Third, older parents, perhaps especially the wife's parents (Daly & Perry, 2017), produced a surplus that could subsidize their younger adult children's children (the grandparent hypothesis) because the older parents' one or two dependent older children largely supported themselves or perhaps even produced a surplus, or their children had all dispersed (Figure 6). Although the simulations did not explicitly model energy transfers between families, food sharing and need-based transfers have been documented in many small-scale societies (Lightner et al., 2023; Ringen et al., 2019; Smith et al., 2019, and references therein).

## 4.1 Rethinking the ECM

The simulation results are broadly consistent with the ECM and the "two-sex" model, albeit with three important distinctions. First, in the two-sex model, although the increasing energetic demands of juveniles play a role, menopause evolves primarily because at midlife the increasing costs of pregnancy and childbirth reduce the benefits of continued reproduction, which then no longer outweigh the benefits of transferring resources to kin. In the simulations, in contrast, menopause evolves simply because the relatively rapid production of offspring who require substantial calories, yet who cannot support themselves for a decade or more, limits the number of dependents to about three or four. With high infant mortality, this limit is reached about 20 years after age at first birth. Nevertheless, age-related increases in the costs of pregnancy and childbirth could be additional factors favoring menopause.

Second, although I have framed the simulations in terms of nuclear families comprising a wife and husband in a lifelong monogamous relationship raising their joint offspring — a key assumption of the two-sex model — the results of the simulations do not rely on this assumption. The results are equally consistent with, e.g., a pooled energy model (Kramer & Ellison, 2010) in which all females reproduce (i.e., no non-reproductive females) and all individuals pool their resources to provision every member of the community according to their needs. In this regard, results are also consistent with the venerable grandmother hypothesis (Hawkes, 2020, 2025), a point to which I will return.

Third, juvenile productivity plays little role in the two-sex model but an essential role in the simulations, where, in combination with reproductive cessation, the increasing productivity of older juveniles (Pretelli et al., 2024, 2022) helps families escape an energy deficit (Figure 6). Without juvenile production, the population would be in negative energy balance (Figure 4). Nevertheless, in mobile hunter-gatherers ranging over a large, spatially and temporally varying landscape replete with competitors, predators, and other sundry hazards (e.g., Blurton Jones et al., 1994), child productivity would have depended on ancestral foraging bands moving to the right place at the right time (Bettinger & Grote, 2016; Hamilton et al., 2016), a critical skill.

Embodied capital should therefore be conceptualized not only as the skills required to extract energy from a resource patch, but also as including the skills required to locate optimal resource patches. Older women and men would have had greater knowledge of the large ranges necessary to support the hunting and gathering lifestyle of a large-bodied hominin (Antón, 2002), and hence greater ability to lead their families to resource patches where high juvenile productivity would have been feasible. This perspective aligns with theories of menopause in toothed whales that emphasize the superior ecological knowledge of older matriarchs (Brent et al., 2015; Rendell et al., 2019), and also with the predominance of female-biased leadership of collective movements in social mammals (Smith et al., 2022). It is also consistent with the anthropology of older women in traditional societies who increasingly take on important social roles (Brown, 1982; Jang et al., 2025), with research on the important ecological knowledge of older individuals in foraging societies (e.g., Biesele & Howell, 1981; Scalise Sugiyama, 2011; Wiessner, 2014; Wood et al., 2024), and with the computational services model, which emphasizes the cognitive challenges of raising multiple dependent offspring (Hagen et al., 2025).

## 4.2 The grandmother hypothesis vs. the ECM “two-sex” hypothesis

The grandmother hypothesis and the ECM both emphasize the importance of food sharing from older individuals with surpluses to younger individuals in deficit, a view strongly supported by simulation results. The primary difference between the hypotheses is the degree of conflict over the distribution of the surpluses. According to the ECM, monogamous couples prefer to transfer their surpluses to their joint children and grandchildren. According to the grandmother hypothesis, in contrast, there is a conflict of interest, with postmenopausal women preferring to transfer surpluses to their adult children and grandchildren, and men preferring to transfer surpluses to fertile women to outcompete other males for mating opportunities. High male intrasexual competition is a consequence of midlife reproductive cessation by females but not males, increasing the pool of fertile males relative to fertile females (Figure S 5) (yet the rarer sex typically has greater bargaining power, and male-biased sex ratios in humans tend to be associated with male commitment to one partner, Schacht et al., 2022).

Because every individual has exactly one biological mother and one biological father, it was convenient to account for the energy production and consumption of the entire population in terms of biological parents and their offspring, i.e., nuclear families. Little hangs on this framing, however, because individuals of the same age and sex had identical energy production and consumption, mortality, IBI's, and ALB's. An individual whose production was less than their consumption required energy transfers (food sharing) from *someone* producing a surplus, who then had less to transfer to others. The simulation models do not address potential conflicts of interest over energy transfers, however, nor do they address the collective action problems or other evolutionary challenges they pose, and therefore they cannot adjudicate between the ECM, the grandmother hypothesis, and other models of food sharing, such as energy pooling (Kramer & Ellison, 2010).

## 4.3 Rethinking the father absent hypothesis to include mother absence

Kuhle (2007) proposed that menopause evolved to mitigate the negative consequences of higher paternal than maternal mortality. With no marriage age gap, only one parent has survived by age 50, on average, and it is almost as likely to be the husband as the wife (Figure S 14). With ALB = 38, the surviving single parent would be caring for about 1.21 older children, whereas with ALB = 60 it would be 2.61 children, some quite young (Figure S 13). As marriage age gaps increase, the age at which only one parent remains decreases, and it is more likely to be the wife, but not dramatically so (Figure S 14). It is possible that limiting the childcare burden on a surviving parent to a few older dependent offspring, which probably would have been manageable, rather than the care of several younger ones, which probably would not have been manageable, was a selection pressure for menopause. However, because the death of the wife would also have been common, and because the fitnesses of monogamous parents are tightly coupled, it is *parental* absence, not necessarily father absence, that matters.[2]

## 4.4 Energetically feasible social organization

In the real world of ancestral foragers, food sharing required physical proximity. The substantial literature on various aspects of hunter-gatherer social organization, which can include patrilocality, matrilocality, bilocality, and neolocality, has ascribed these patterns to, e.g., inbreeding avoidance, male coalitions, paternity uncertainty, ecological variation, childcare, and warfare (e.g.,

[2] As an aside, the not infrequent death of mothers with multiple dependent children could have helped select for female fidelity to ensure surviving fathers would continue to invest in their joint offspring.

Barnard, 1983; Ember, 1975; Hamilton et al., 2016, 2007; Kelly, 2013; Marlowe, 2004; Moravec et al., 2018). Simulation results suggest that because age- and sex-based variation in fertility, mortality, and food acquisition rates in particular local ecologies determined energy consumption and production, thus determining energetically feasible group compositions, these factors probably played pivotal roles shaping hunter-gatherer social organization.

# 5 Limitations

Results depended on strong assumptions, such as population stationarity (constant age structure and no growth). If ancestral human populations were instead characterized by, e.g., fast growth with frequent population crashes (Gurven & Davison, 2019), this would alter the number of individuals requiring energy subsidies relative to the number able to provide them. Nevertheless, energy production and consumption across the lifespan of a family with the given age structure would remain valid. Results also depended on the assumptions that all women reproduced at a constant IBI starting at age 20 until ALB, and that age- and sex-specific productivity did not respond to individual and family energy needs. Alternatively, physiological mechanisms linking fertility to energy metabolism (Fontana & Della Torre, 2016; Torre et al., 2014) and breastfeeding duration (Calik-Ksepka et al., 2022), and behavioral mechanisms like postpartum sex taboos (Mchome et al., 2020), might have adjusted AFB and/or IBI, and thus offspring energy demand, in response to the availability of energy in the environment. Moreover, parents and children could have adjusted the time devoted to production in response to changes in demand (e.g., Kramer, 2005). If so, these would mitigate the negative impact of a later ALB on energy balance. Future studies should investigate which strategies for matching energy production to consumption across the lifecourse optimized fitness.

In addition, parameters were varied independently (with the exception of male and female productivity, whose sum was constrained), yet in ancestral foragers, skill acquisition rates, productivity, the importance of strength, interbirth intervals, marriage age gaps, and so forth, likely co-varied in complex ways, and were linked to socioecological variation. The multidimensional parameter space used in the simulations did not incorporate the unknown covariation of these values across actual foraging societies, past or present, and averages across the parameter space do not represent an "average" ancestral foraging society.

Finally, insights into a possible energetic selection pressure for an earlier rather than later ALB do not rule out byproduct or other adaptive hypotheses for menopause.

# 6 Concluding remarks

Simulations revealed a human-specific midlife energetic constraint on continued production of slowly developing, energetically expensive offspring that, perhaps in conjunction with other constraints such as mammalian reproductive senescence (Huber & Fieder, 2018) and kin competition in hominins (White et al., 2025), might help explain human's distinctive midlife reproductive cessation and high levels of cooperative food sharing.

## 7 Acknowledgments

Thanks to Ray Hames, Nicole Hess, Ilaria Pretelli, Erik Ringen, and Eric Schniter for helpful comments.

# 9 Supplementary information

## 9.1 Hunter-gatherer demography and life history

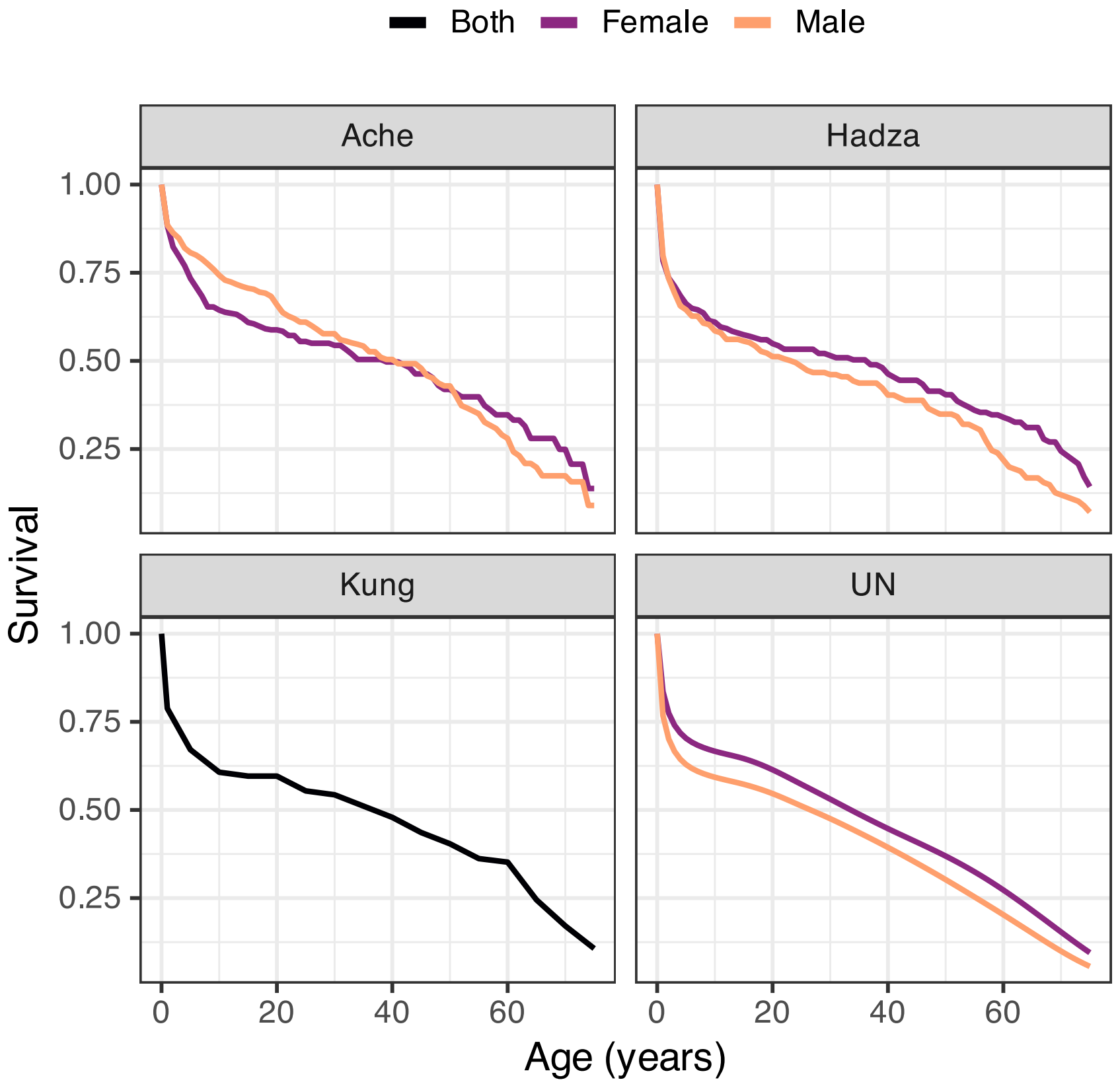

Figure S 1: Survival curves. Ache data from Hill & Hurtado (1996). Hadza data from Blurton Jones (2016), !Kung data from Howell (2017). UN values for female $e_0 = 35$ and male $e_0 = 30$ from UN life tables (Li & Gerland, 2012), using data and code from Gaddy et al. (2025).

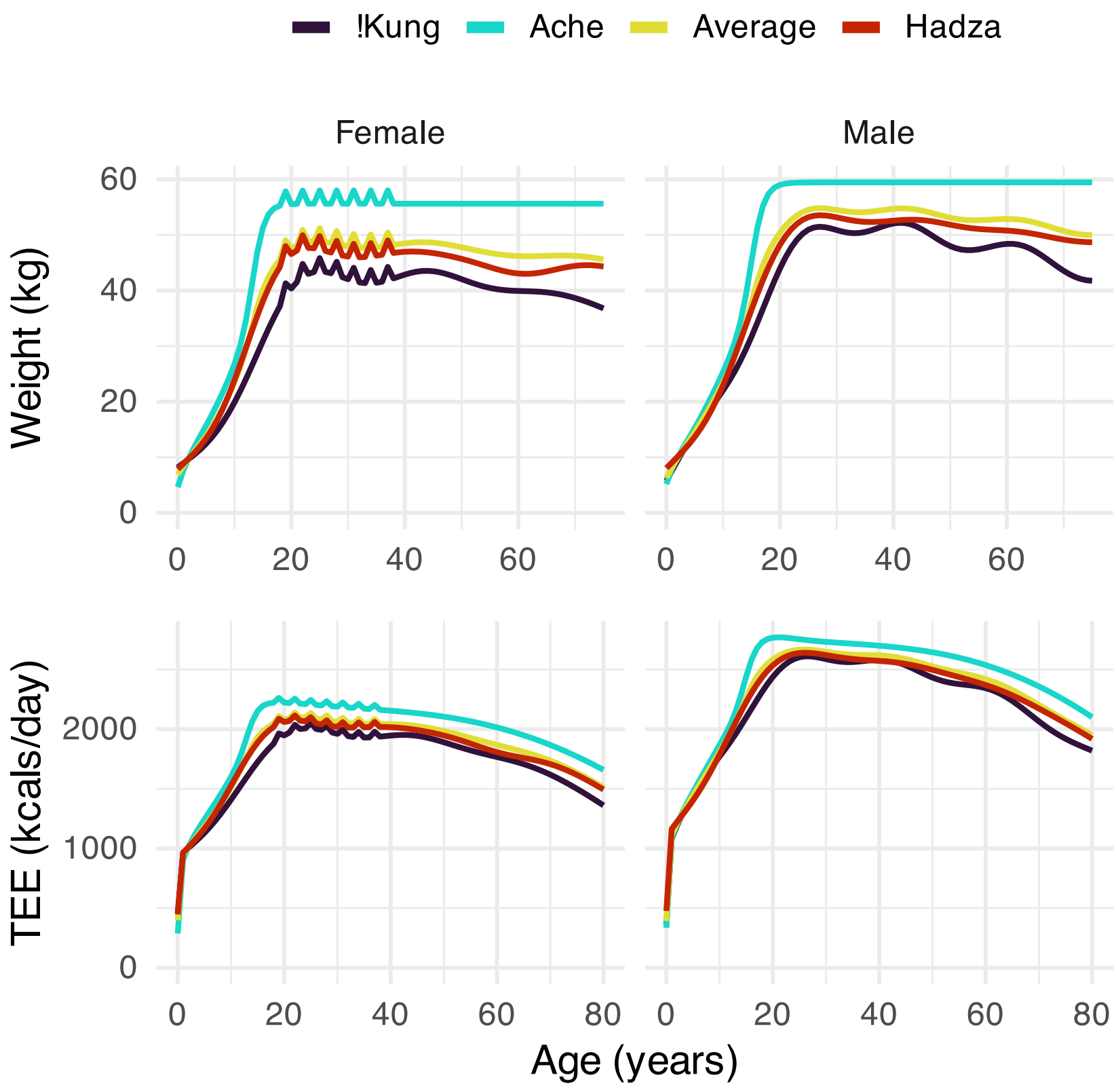

Figure S 2: Weight by sex and age (top) and total energy expenditure (TEE) by age and sex (bottom). Female weight and TEE includes pregnancy increase. Kung weight data from Howell (2009). Ache weight data from Walker et al. (2005). Hadza weight data from Blurton Jones (2016). TEE values from the equation in Bajunaid et al. (2025), except values for age 0-1 from Pontzer et al. (2021).

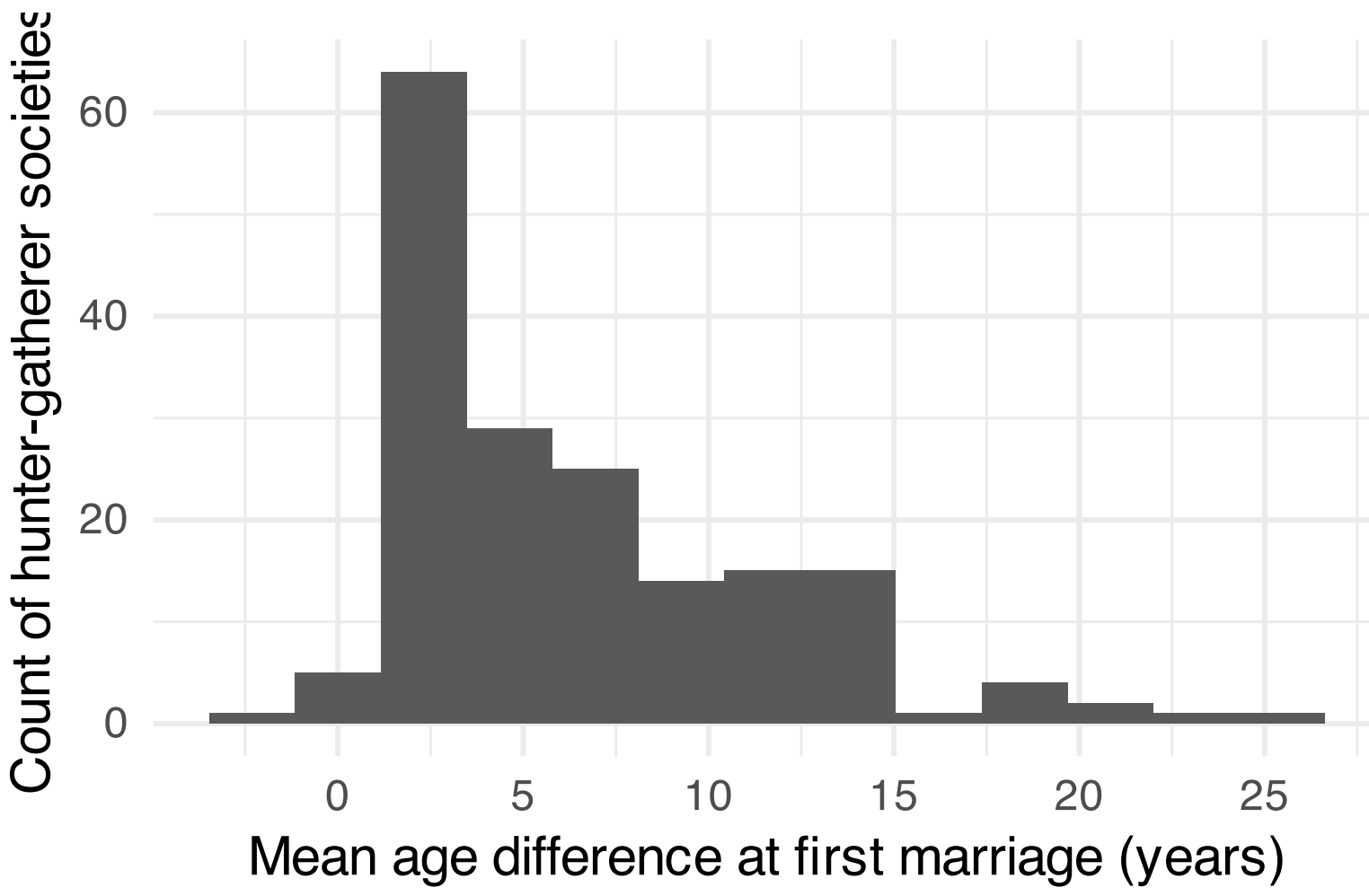

Figure S 3: Histogram of the mean differences in age at first marriage for husbands and wives in 177 hunter-gatherer societies (positive values indicate older ages for men). Note that these values do not necessarily represent the age difference between spouses because young men's and women's first marriages could conceivably have been to much older individuals (which would not be the first marriages for those much older individuals). Data from Marwick et al. (2016).

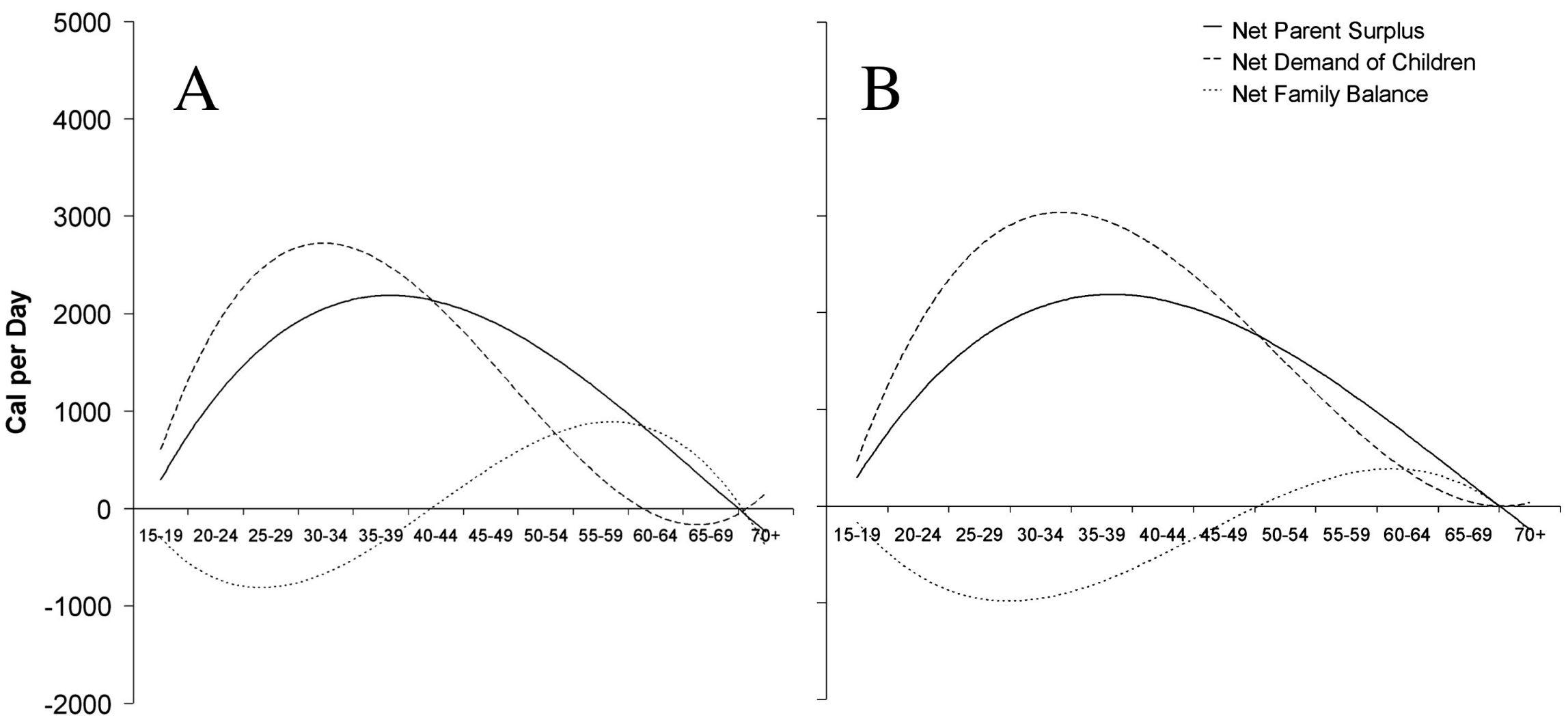

Figure S 4: **A**: Tsimane parental production, children's demands, and net family production. **B**: Simulated values based on delayed menopause, showing an increase in energy deficits and a decrease in energy surpluses. Figure from Kaplan et al. (2010).

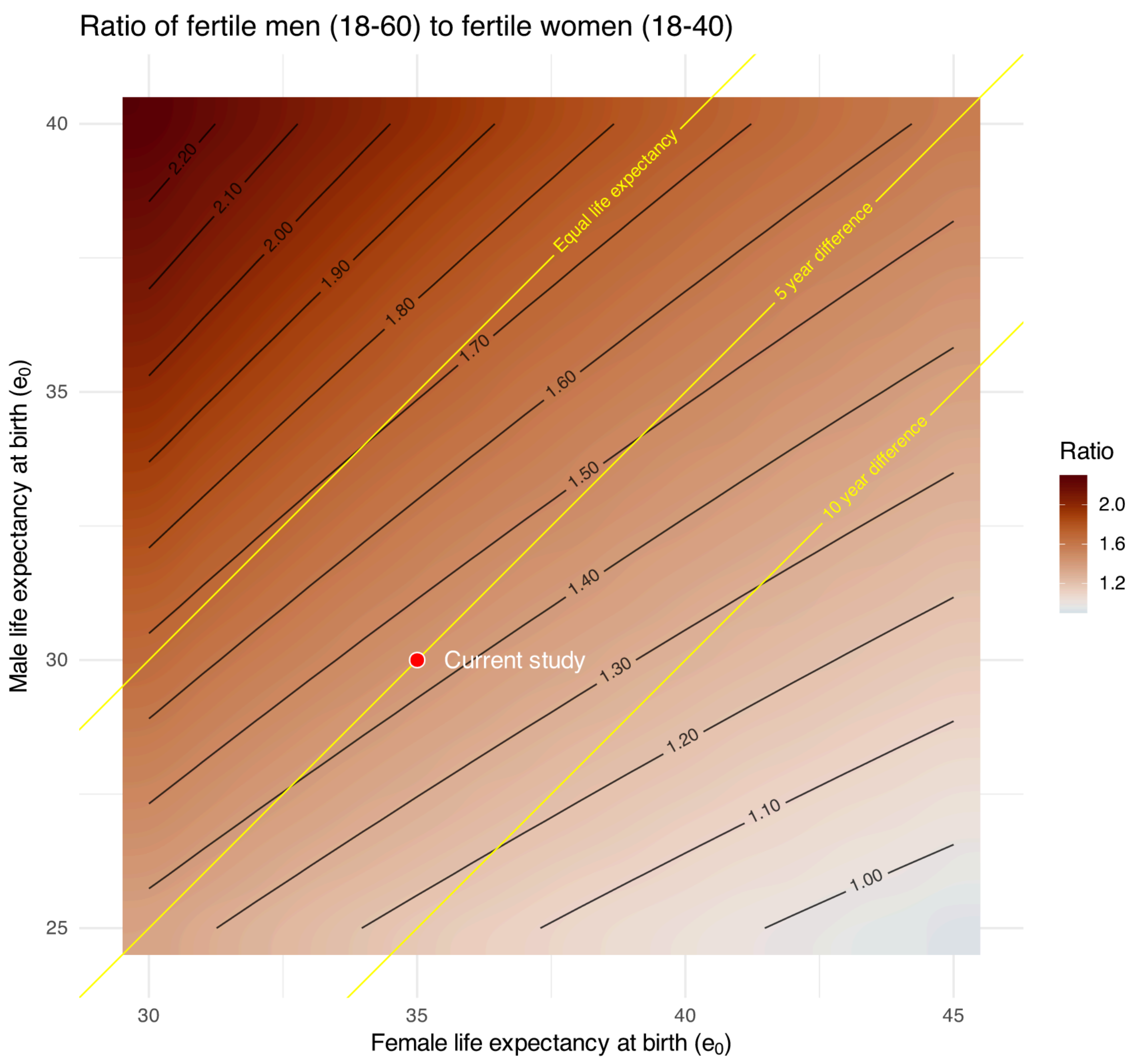

Figure S 5: Ratio of fertile men (18-60) to fertile women (18-40) by female and male life expectancies at birth in a stationary population. Color gradient and contour lines represent the ratios for different female and male life expectancies. The red dot indicates the life expectancies used in the current study, and the associated ratio of about 1.44 adult men for each fertile woman, which implies male intrasexual competition for mates, a central feature of the grandmother hypothesis.

## 9.2 Effects of variation in skill ontogeny parameters on energy balance

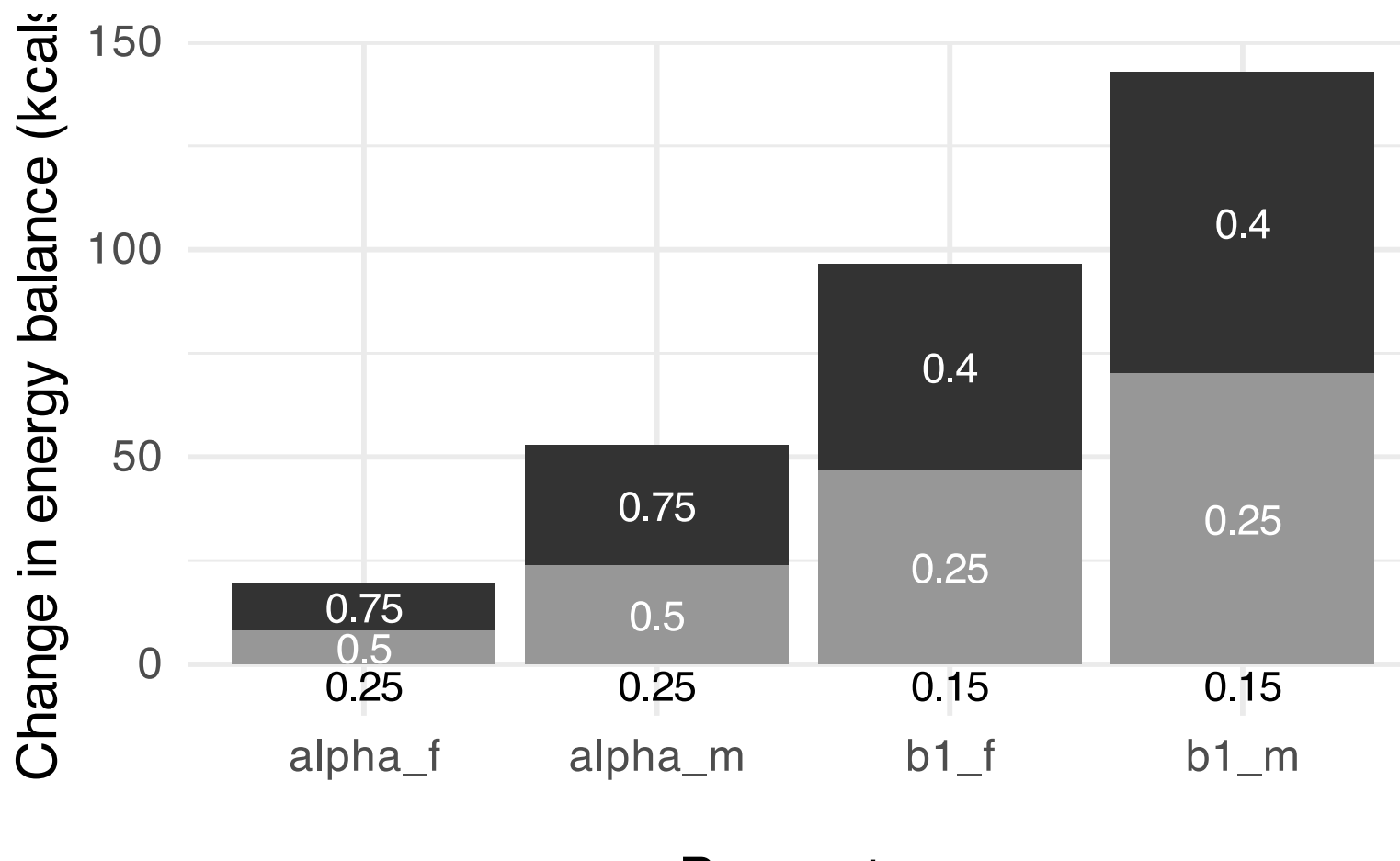

Figure S 6: Mean difference in energy balance for each value of each skill ontogeny parameter, relative to its next smallest value, for individuals ages 0-40 in the population model (averaged over the values of the other parameters). Values on the bars, and at the base of the bars, are the parameter values.

## 9.3 Ratio of child production to child consumption

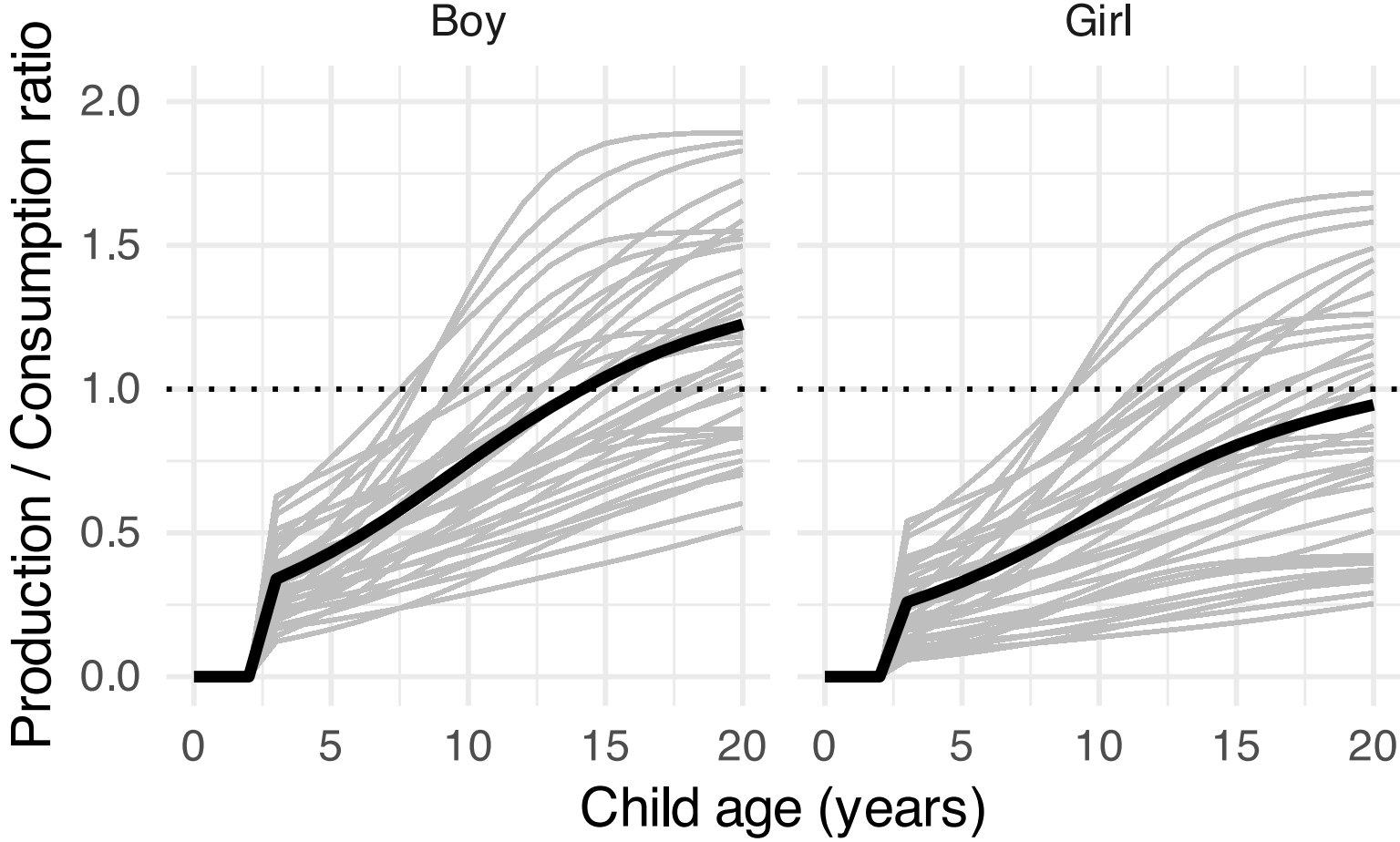

Figure S 7: The ratios of individual children's production to their consumption by age. Grey lines represent trajectories for different combinations of parameter values. Black lines represent the mean ratios averaged across the parameter space.

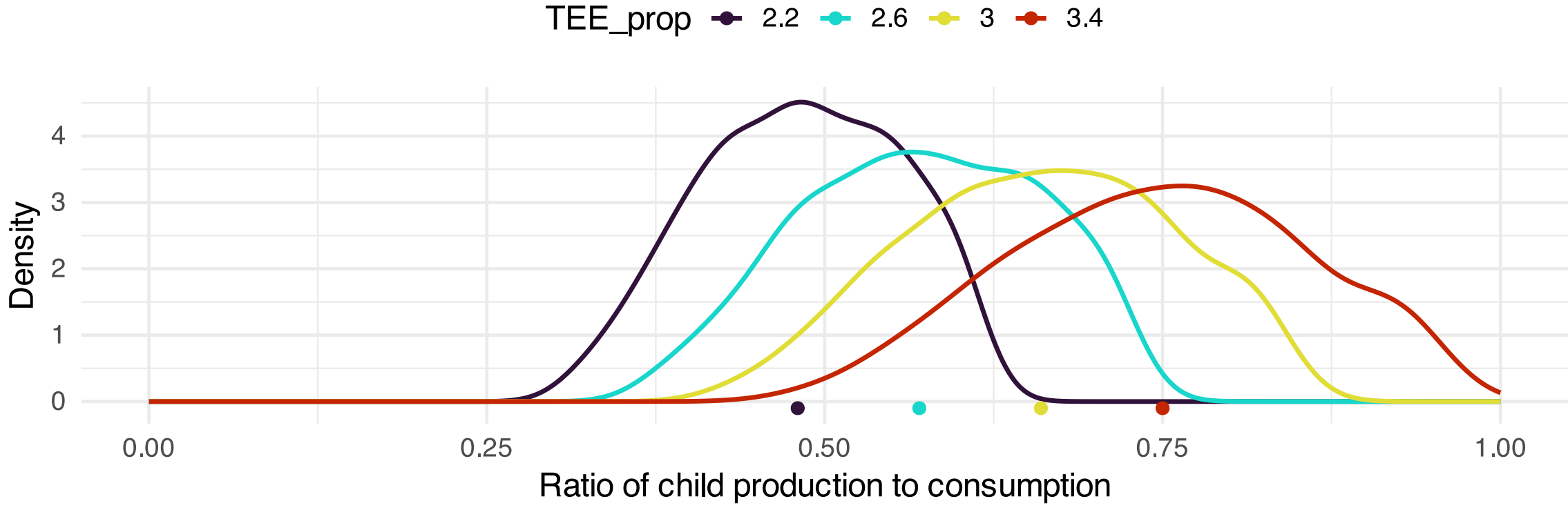

Figure S 8: Distributions of the ratios of the total production of all dependent children (age < 20) to their total consumption, across the parameter space, for different values of $TEE_{prop}$. Dots along the x-axis are the mean values of each distribution.

## 9.4 Effects of variation in ALB and IBI on family composition and energy consumption

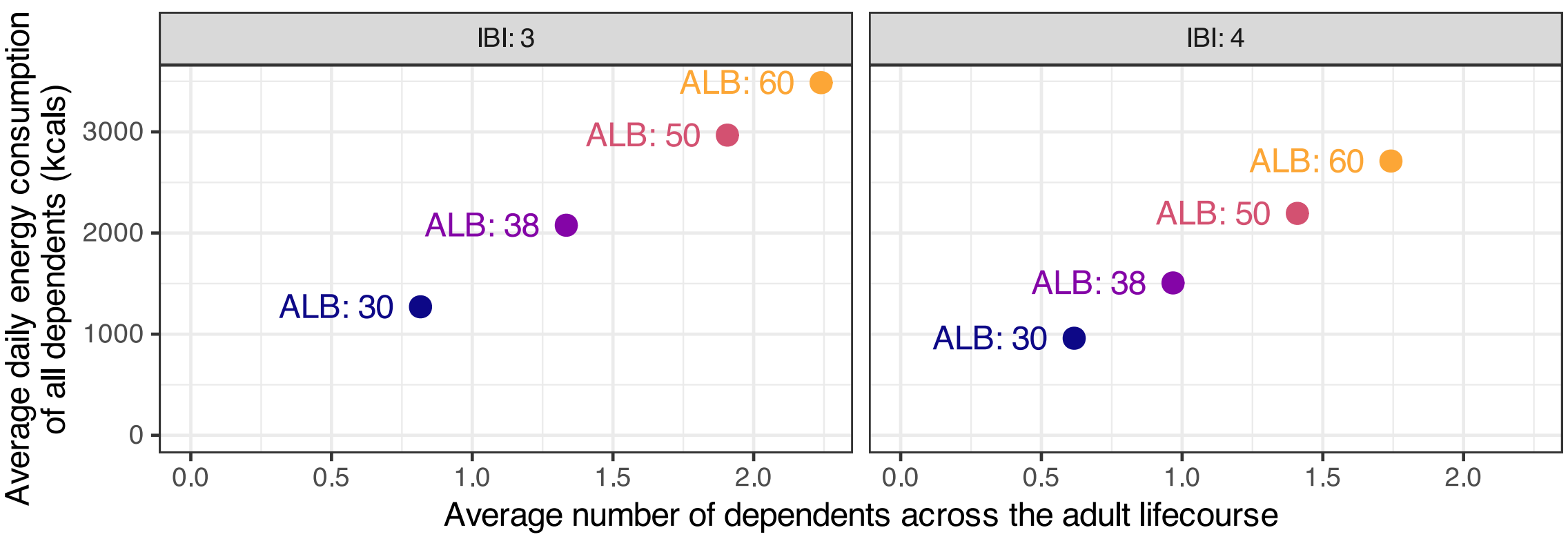

Figure S 9: Average number and daily energy consumption of dependents that families were supporting each day across the adult lifespan, for each value of the ALB and IBI.

## 9.5 Energy balance distributions

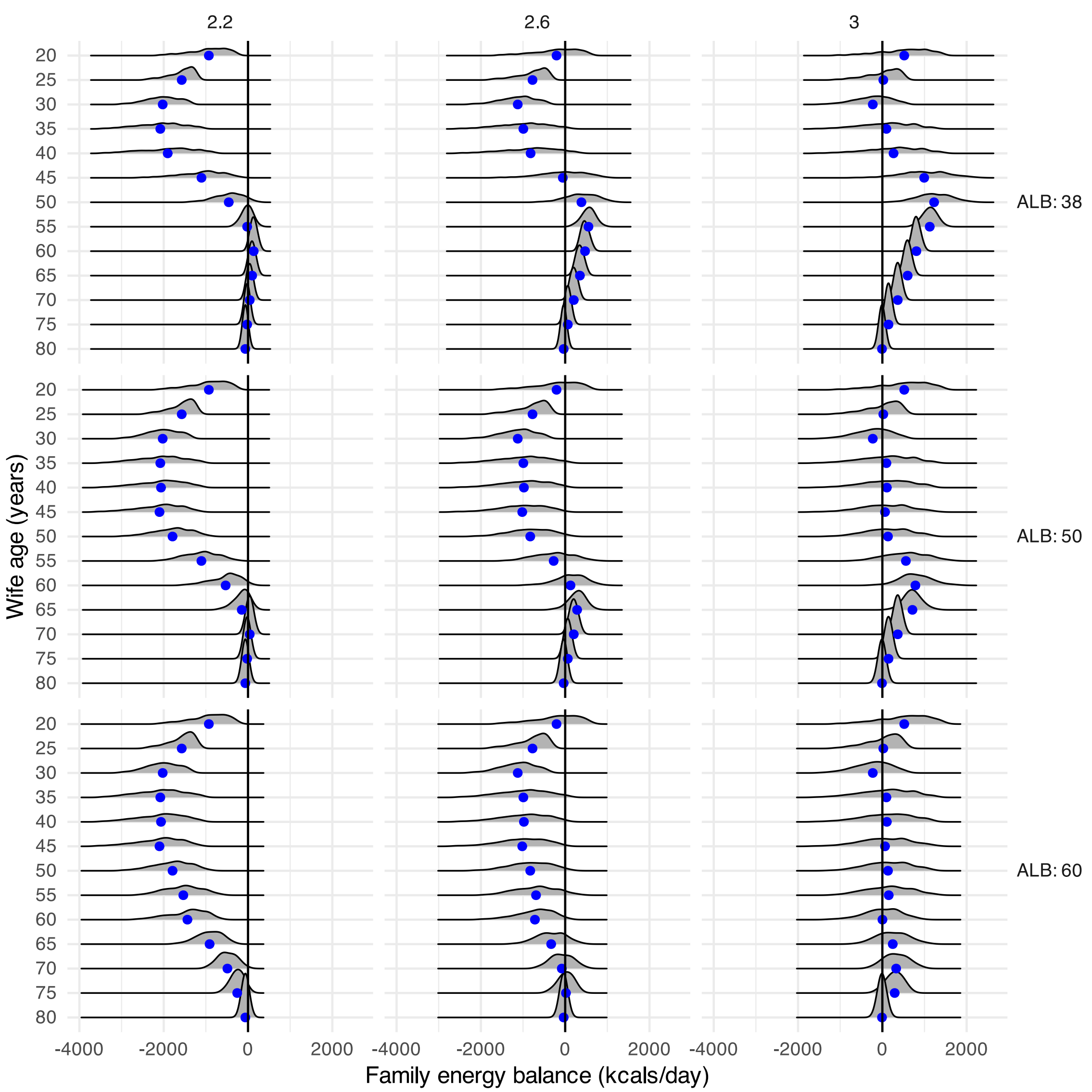

Figure S 10: Variation in energy balance at each age due to variation in skill ontogeny parameters and IBI. Columns: joint adult productivity as a proportion of adult TEE ($TEE_{prop,f}$ + $TEE_{prop,m}$). Rows: ages of last birth (ALB). Blue dots are the mean values. Joint productivity and ALB restricted to intermediate values for clarity.

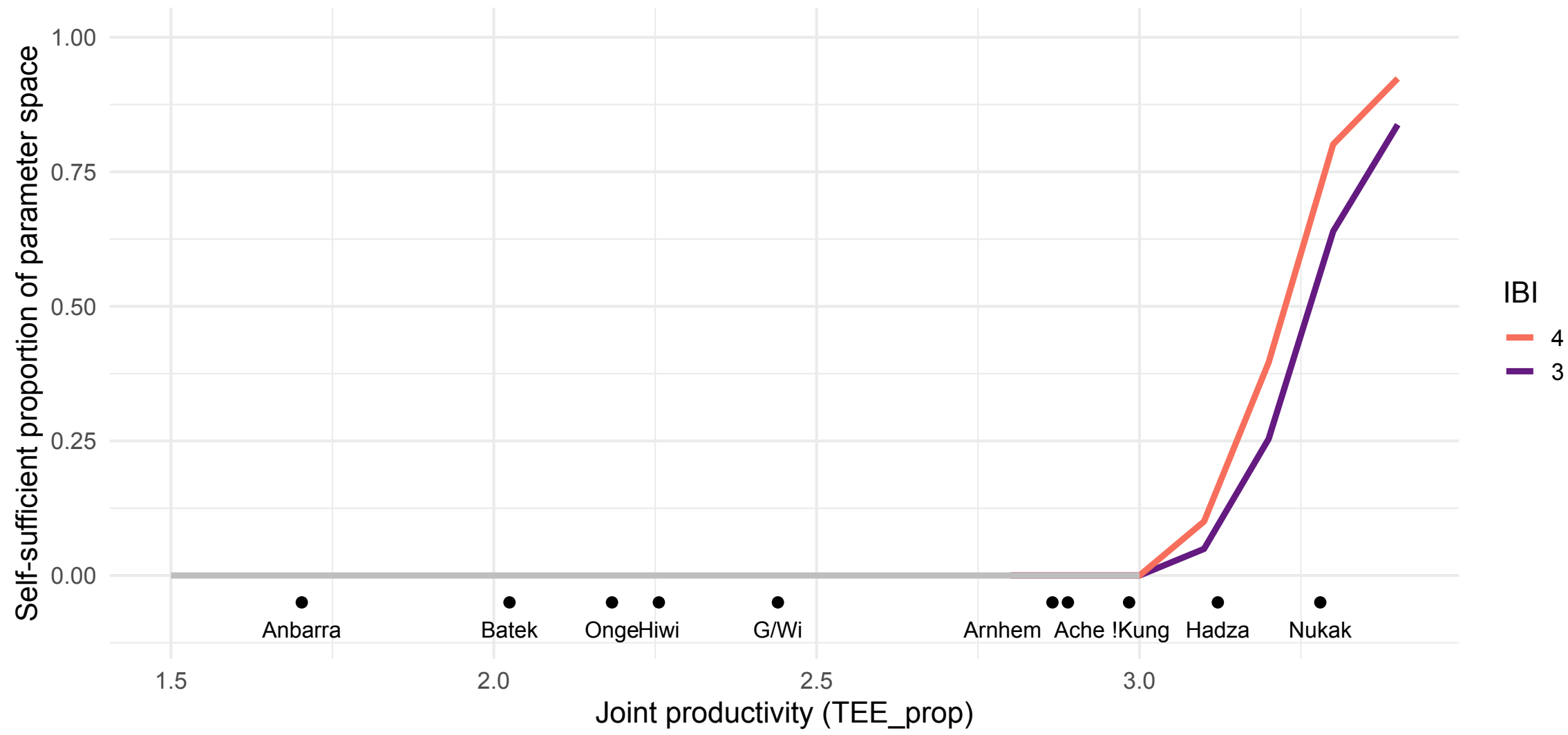

Figure S 11: The proportion of the parameter space in which family energy balance never falls below 0, and families therefore do not need energy transfers from others, as a function of the joint production of the wife and husband. Dots along the x-axis are the joint energy production values from contemporary hunter-gatherers in Figure 3, for comparison.

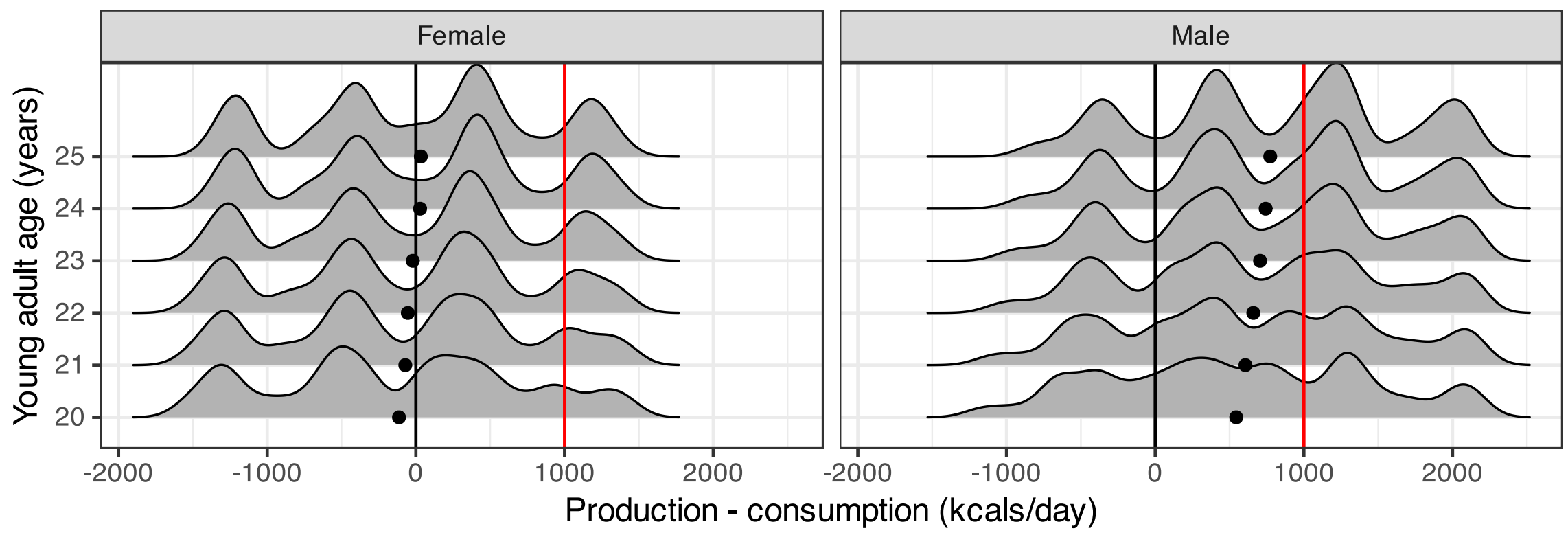

Figure S 12: Variation in net energy production of young adults due to variation in skill ontogeny parameters. The black dots represent the means of each distribution. The vertical red line represents the approximate TEE of an infant or very young child. The four modes are caused by the four sex-specific values of $TEE_{prop}$. Young adults' positive net energy production could help provision young siblings (helper-at-the-nest), or the siblings of one's spouse, e.g., brideservice.

## 9.6 Dependents and surviving parents

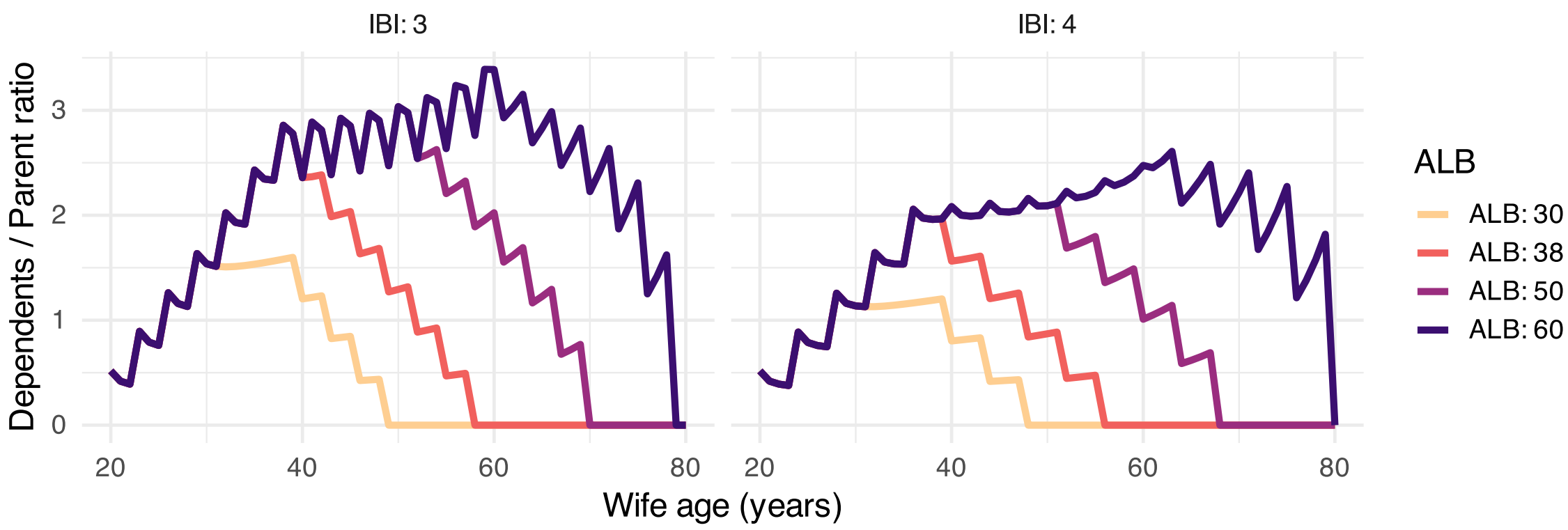

Figure S 13: Ratio of the number of dependents to the number of surviving parents for varying values of ALB, IBI, and wife's age.

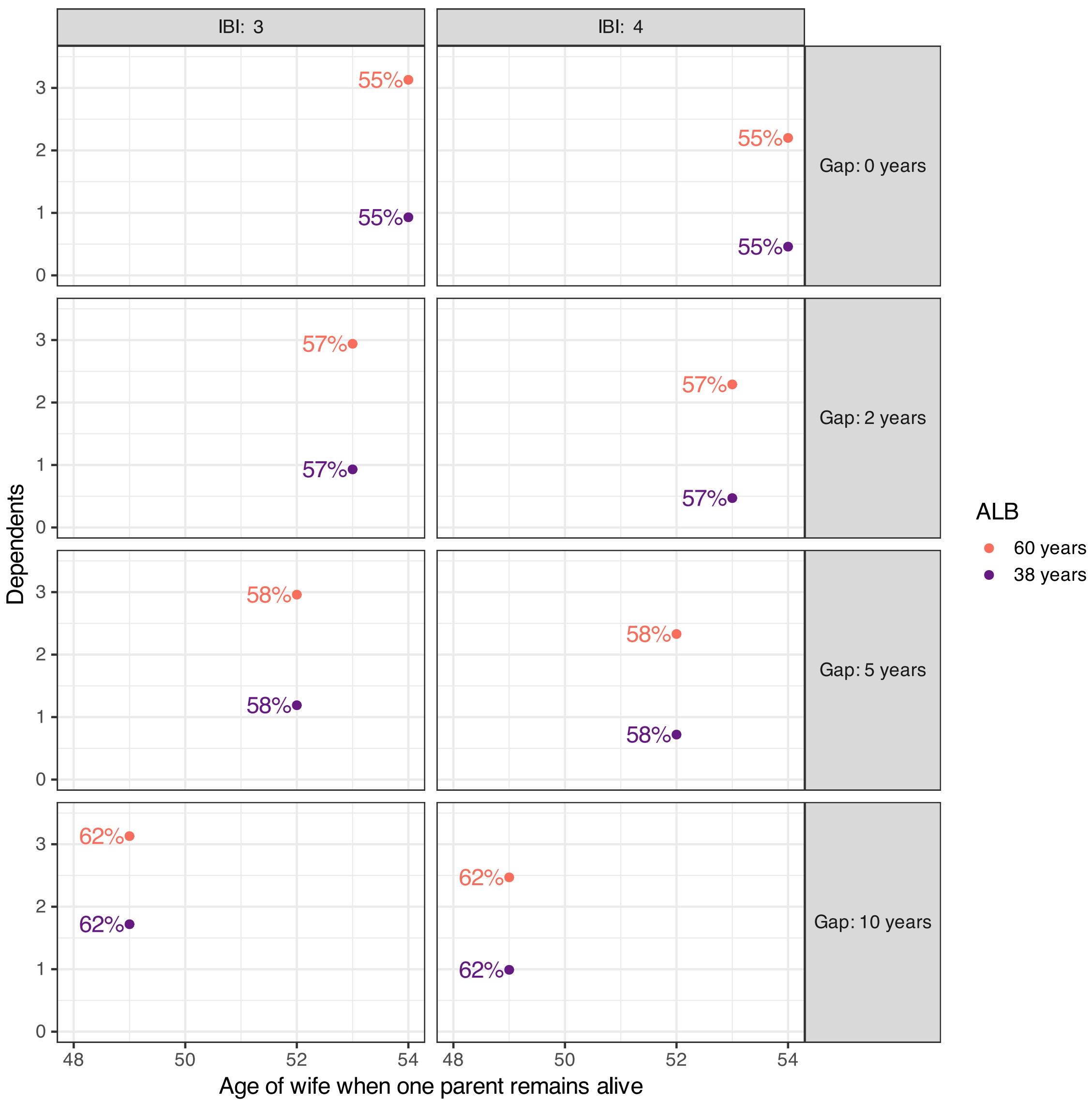

Figure S 14: The number of dependents, on average, at the age when only one parent remains alive due to wife and husband mortality. Gap is the age gap in marriage. Percentage values are the probabilities that the one remaining parent is the wife.